%% file: sample-acmtog-SIGGRAPH-submission.tex
\documentclass[sigconf]{acmart}

\usepackage[commandnameprefix=always]{changes}

\usepackage{booktabs} 

\usepackage{pgffor}

\usepackage{placeins}
\usepackage{subfig}

\usepackage[ruled]{algorithm2e} 

\SetAlFnt{\small}
\SetAlCapFnt{\small}
\SetAlCapNameFnt{\small}
\SetAlCapHSkip{0pt}

\copyrightyear{2024}
\acmYear{2024}
\setcopyright{rightsretained}
\acmConference[SA Conference Papers '24]{SIGGRAPH Asia 2024 Conference Papers}{December 3--6, 2024}{Tokyo, Japan}
\acmBooktitle{SIGGRAPH Asia 2024 Conference Papers (SA Conference Papers '24), December 3--6, 2024, Tokyo, Japan}
\acmDOI{10.1145/3680528.3687693}
\acmISBN{979-8-4007-1131-2/24/12}

\begin{document}
\title{Polar Interpolants for Thin-Shell Microstructure Homogenization}

\author{Antoine Chan-Lock}
\orcid{0000-0002-4632-9579}
\affiliation{%
 \institution{Universidad Rey Juan Carlos}
 \city{Madrid}
 \country{Spain}}
\email{antoine.chanlock@gmail.com}

\author{Miguel A. Otaduy}
\orcid{0000-0002-3880-7622}
\affiliation{%
 \institution{Universidad Rey Juan Carlos}
 \city{Madrid}
 \country{Spain}}
\email{miguel.otaduy@urjc.es}



\begin{abstract}
This paper introduces a new formulation for material homogenization of thin-shell microstructures.
It addresses important challenges that limit the quality of previous approaches: methods that fit the energy response neglect visual impact, methods that fit the stress response are not conservative, and all of them are limited to a low-dimensional interplay between deformation modes.
The new formulation is rooted on the following design principles: the material energy functions are conservative by definition, they are formulated on the high-dimensional membrane and bending domain to capture the complex interplay of the different deformation modes, the material function domain is maximally aligned with the training data, and the material parameters and the optimization are formulated on stress instead of energy for better correlation with visual impact.
The key novelty of our formulation is a new type of high-order RBF interpolant for polar coordinates, which allows us to fulfill all the design principles.
We design a material function using this novel interpolant, as well as an overall homogenization workflow.
Our results demonstrate very accurate fitting of diverse microstructure behaviors, both quantitatively and qualitatively superior to previous work.
\end{abstract}

%
%


%
%


\begin{teaserfigure}
  \includegraphics[width=0.32\textwidth]{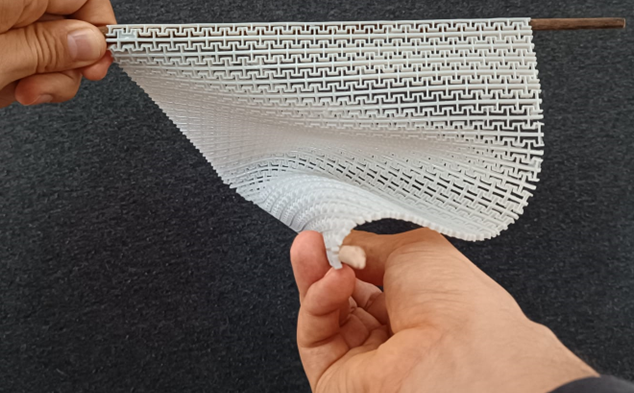}
  \includegraphics[width=0.33\textwidth]{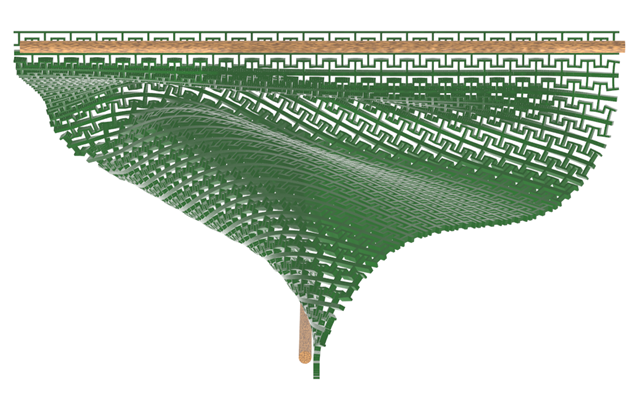}
  \includegraphics[width=0.33\textwidth]{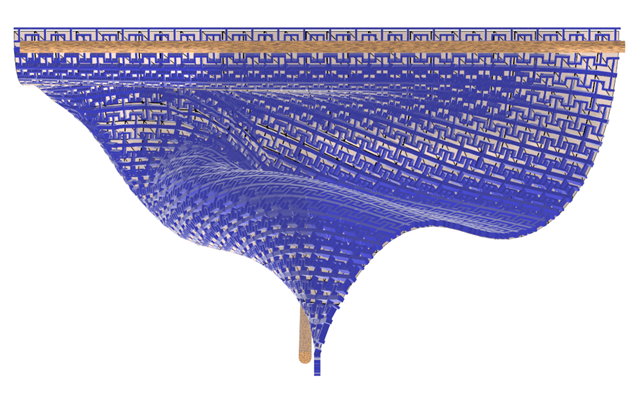}
  \caption{The images compare a twist deformation (far from the training data) of a real-world thin-shell micro-texture (left), a micro-scale simulation (center), and a macro-scale simulation using our homogenized material model (right).} 
  \Description{pass}
  \label{fig:teaser}
\end{teaserfigure}


\maketitle


\input{notations0}

\input{includes/intro}
\input{includes/related}
\input{includes/thinshell}
\input{includes/interpolants}
\input{includes/homogenization}
\input{includes/experiments}
\input{includes/conclusion}
\clearpage

\bibliographystyle{ACM-Reference-Format}
\bibliography{refs}

\input{includes/validations}


\end{document}

%% file: notations0.tex
\let\rm=\rmfamily
\let\sf=\sffamily
\let\tt=\ttfamily
\let\it=\itshape
\let\sl=\slshape
\let\sc=\scshape
\let\bf=\bfseries

\renewcommand{\vec}[1]{\mathbf{#1}}
\newcommand{\mat}[1]{\mathbf{#1}}
\newcommand{\set}[1]{\mathcal{#1}}
\newcommand{\func}[1]{\mathrm{#1}}

\def \N {\mbox{\rm \hbox{I\kern-.15em\hbox{N}}}}
\def \R {\mbox{\rm \hbox{I\kern-.15em\hbox{R}}}}
\def \laplace {\Delta}
\def \grad {\nabla}
\newcommand{\of}[1]{\!\left( #1 \right)}
\newcommand{\abs}[1]{\left| #1 \right|}
\newcommand{\norm}[1]{\left\Vert {#1} \right\Vert}
\renewcommand{\matrix}[2]{\left(\begin{array}{#1}#2\end{array}\right)}
\newcommand{\twovec}[2]{\left(\begin{array}{c}#1\\#2\end{array}\right)}
\newcommand{\threevec}[3]{\left(\begin{array}{c}#1\\#2\\#3\end{array}\right)}
\newcommand{\fourvec}[4]{\left(\begin{array}{c}#1\\#2\\#3\\#4\end{array}\right)}
\newcommand{\DD}[2] {\frac{\partial{#1}}{\partial{#2}}}
\newcommand{\DDTwo}[2] {\frac{\partial^2{#1}}{\partial{#2}^2}}
\newcommand{\DDThree}[2] {\frac{\partial^3{#1}}{\partial{#2}^3}}
\newcommand{\DDfull}[2] {\frac{{\text d}{#1}}{{\text d}{#2}}}
\newcommand{\DDp}[2] {\frac{\partial\left({#1}\right)}{\partial{#2}}}
\newcommand{\DDfullTwo}[2] {\frac{{\text d}^2{#1}}{{\text d}{#2}^2}}
\newcommand{\Sr}[1] {{#1}^*}
\newcommand{\Srp}[1] {\left({#1}\right)^*}
\newcommand{\sk}[1] {{\text{skew}\left({#1}\right)}}

\newcommand{\refapp}[1]{Appendix~\ref{sec:#1}}
\newcommand{\refsec}[1]{Section~\ref{sec:#1}}
\newcommand{\refeq}[1]{(\ref{eq:#1})}
\newcommand{\refeqs}[2]{(\ref{eq:#1})-(\ref{eq:#2})}
\newcommand{\reffig}[1]{Fig.~\ref{fig:#1}}
\newcommand{\reftab}[1]{Table~\ref{tab:#1}}
\newcommand{\cf}[1]{cf.\ Fig.~\ref{fig:#1}}
\newcommand{\eq}[1]{(\ref{eq:#1})}

\newcommand{\Ithree} {{\mat{I}_{3\times3}}}

\newcommand{\cm} {$\backslash\backslash$}

\renewcommand{\labelenumii}{\arabic{enumii}$)$ }
\renewcommand{\labelenumiii}{\arabic{enumiii}$)$ }

\definecolor{amethyst}{rgb}{0.6, 0.4, 0.8}
\definecolor{darkpastelgreen}{rgb}{0.01, 0.75, 0.24}
\newcommand{\Antoine}[1]{\textcolor{darkpastelgreen}{[\textbf{Antoine}: {#1}]}} 
\newcommand{\Miguel}[1]{\textcolor{amethyst}{[\textbf{Miguel}: {#1}]}}

%% file: includes/intro.tex
\section{Introduction}
\label{sec:introduction}

Real-world materials exhibit complex nonlinear and anisotropic behaviors that are difficult or impossible to reproduce with standard constitutive material models.
For this reason, we have seen many attempts at modeling these complex behaviors using strain-space interpolation techniques~\cite{Wang2011,Miguel2012,Xu2015}.
Microstructures are particularly challenging materials.
By designing the micro-scale material distribution, one can achieve macro-scale behaviors that even surpass the nonlinearity and anisotropy of natural materials~\cite{Bickel2010,Schumacher:2015:MCE:2809654.2766926,Panetta2015,Konakovic2018}.

In this work, we study the material homogenization of thin-shell microstructures, i.e., the design of constitutive material models that match the macro-scale membrane (in-plane) and bending (out-of-plane) response of thin-shell microstructures.
Such thin-shell microstructures pose important challenges to material homogenization: the continuum deformation space is six-dimensional, anisotropy arises along arbitrary directions, and there may be a complex interplay between membrane and bending deformations.
For instance, the microstructure shown in \reffig{teaser} releases energy in two complex ways when stretched: it exhibits auxetic behavior (i.e., it widens when stretched), and it produces characteristic folds (See also \reffig{bigtable4}).

When designing interpolation material models for thin shells, we set the following desiderata: they should model the membrane-bending interplay, they should be conservative, and they should optimize the strain-stress response, as this shows good correlation with visual impact.
Despite the large number of previous methods, to the best of our knowledge none of them meets all these criteria.
We provide a detailed discussion in \refsec{prev}, but in summary, previous methods either fit the strain-stress response and are not conservative~\cite{Wang2011}, or they fit the energy response to be conservative but then fail to account for visual impact~\cite{Miguel2016,sperl2020hylc}.
We present {\em polar interpolants}, a novel approach for material homogenization of thin-shell microstructures that meets all desiderata.

The first ingredient of our approach is to represent both membrane and bending strain in terms of magnitude and direction, as detailed in \refsec{thinshell}.
We refer to this representation as {\em eigen-strain}.
This strain representation is not new, and it has several advantages: it is easier to produce deformation data that is well aligned with the function domain, and fitting quality improves because the function domain is less curved.
However, the eigen-strain representation complicates the design of a conservative material model that optimizes the strain-stress relationship.
We are not aware of any suitable interpolation-based material that handles the periodicity of the eigen-strain representation,

The second ingredient of our approach is the design of high-order polar interpolants, i.e., interpolation functions that fit high-order derivatives on periodic domains.
Our polar interpolants are an extension of the high-order interpolants of Chan-Lock et al.~\shortcite{ChanLock2022} to periodic domains.
In \refsec{interp} we detail their formulation and we show how we design interpolation-based thin-shell energy models as a function of eigen-strain.

The third and last ingredient of our approach is the workflow for material homogenization, which we describe in \refsec{homogenization}.
We present the generation of training data and the progressive refinement of the material model.

With all ingredients together, we achieve high-quality fitting of complex thin-shell microstructures, both qualitatively and quantitatively superior to previous work~\cite{sperl2020hylc}.
Even though our training data is made of uniform deformations with uniaxial curvature only, we show good match of non-uniform deformations with biaxial curvature, which extend well beyond the training conditions.
We also show qualitative matching of real-world 3D-printed microstructures.

All the data and code of our work can be found in the following URL
\href{https://github.com/antoine-chan-lock/PolarInterpolants}{https://github.com/antoine-chan-lock/PolarInterpolants}.

%% file: includes/related.tex
\section{Related Work}
\label{sec:prev}

\begin{figure}[t!]
    \centering
    \includegraphics[trim={0cm 4cm 0cm 4cm},clip,width=.75\linewidth]{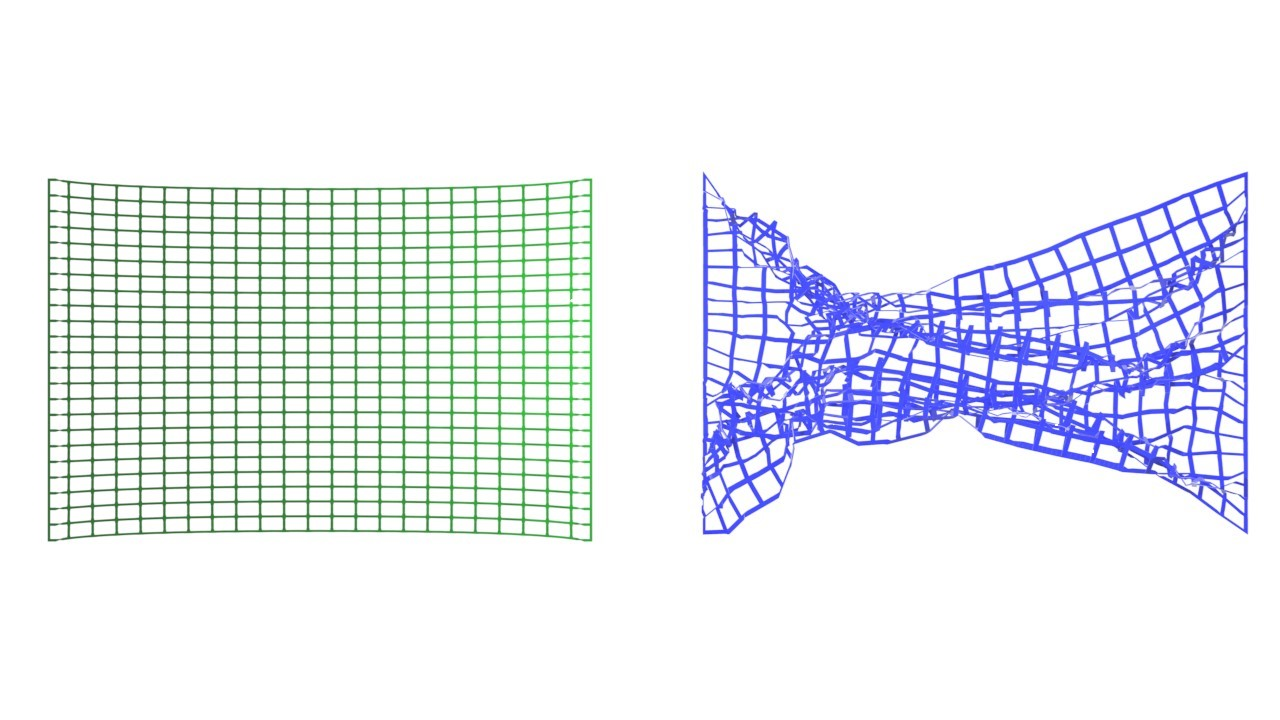}
    \caption{Conservativeness of the mechanical model is critical for simulation robustness.
    The same microstructure is simulated with energy-based line-search optimization (left), suitable for conservative models, and with stress-norm line-search optimization (right), the common choice for non-conservative models.
    The latter fails to converge.}
\Description[Motivation for model conservativeness]{Motivation for model conservativeness.}
\label{fig:Conservativeness}
\end{figure}

The most popular approach to model a material's mechanical response is to use a constitutive law with few parameters ~\cite{Li2015,Smith2018,Kim2020,Wen2023}.
Consequently, many works have addressed the characterization of thin-shell materials (e.g., cloth) using low-dimensional parametric models.
Bhat et al.~\shortcite{Bhat2003} pioneered the estimation of cloth material parameters from video.
In recent years, efforts are aimed at simplifying the characterization methodology, in particular by leveraging large data sets for learning-based material estimation from limited data~\cite{Feng2022,Rodriguez2023}.

Using low-dimensional parametric models works well for authoring VFX or when an average fit to the mechanical response is sufficient.
However, research on microstructures soon showed that these low-dimensional parametric models cannot capture the richness of microstructure behaviors, and researchers moved to strain-dependent material parameterizations~\cite{Bickel2010}.

When choosing a strain-dependent material parameterization, one of the properties often desired is conservativeness, as it can guarantee simulation stability and robustness via line-search energy optimization~\cite{Gast2015,Li2020}.
In the absence of conservativeness, a common choice for line search is the stress norm, but this is not robust for our microstructure simulations, as shown in \reffig{Conservativeness}.
Sperl et al.~\shortcite{sperl2020hylc} achieved conservativeness by modeling the material as an energy function and fitting measured energy values.
However, we have observed that, for thin-shell microstructures, energy shows little correlation with visually apparent deformation.
\reffig{EnergyMotivation} compares the same microstructure under stretch and bending.
Due to the large difference in stiffness, the energy is $100\times$ higher in the stretch deformation.
A least-squares fitting of energy values would dramatically blur the response under bending.

\newcommand{\motivation}[1]{
\includegraphics[trim={20cm 12cm 20cm 10cm},clip,width=.31\linewidth]{images/EnergyMotivation/output1_#1.png}
}

\begin{figure}[t!]
    \centering
    \motivation{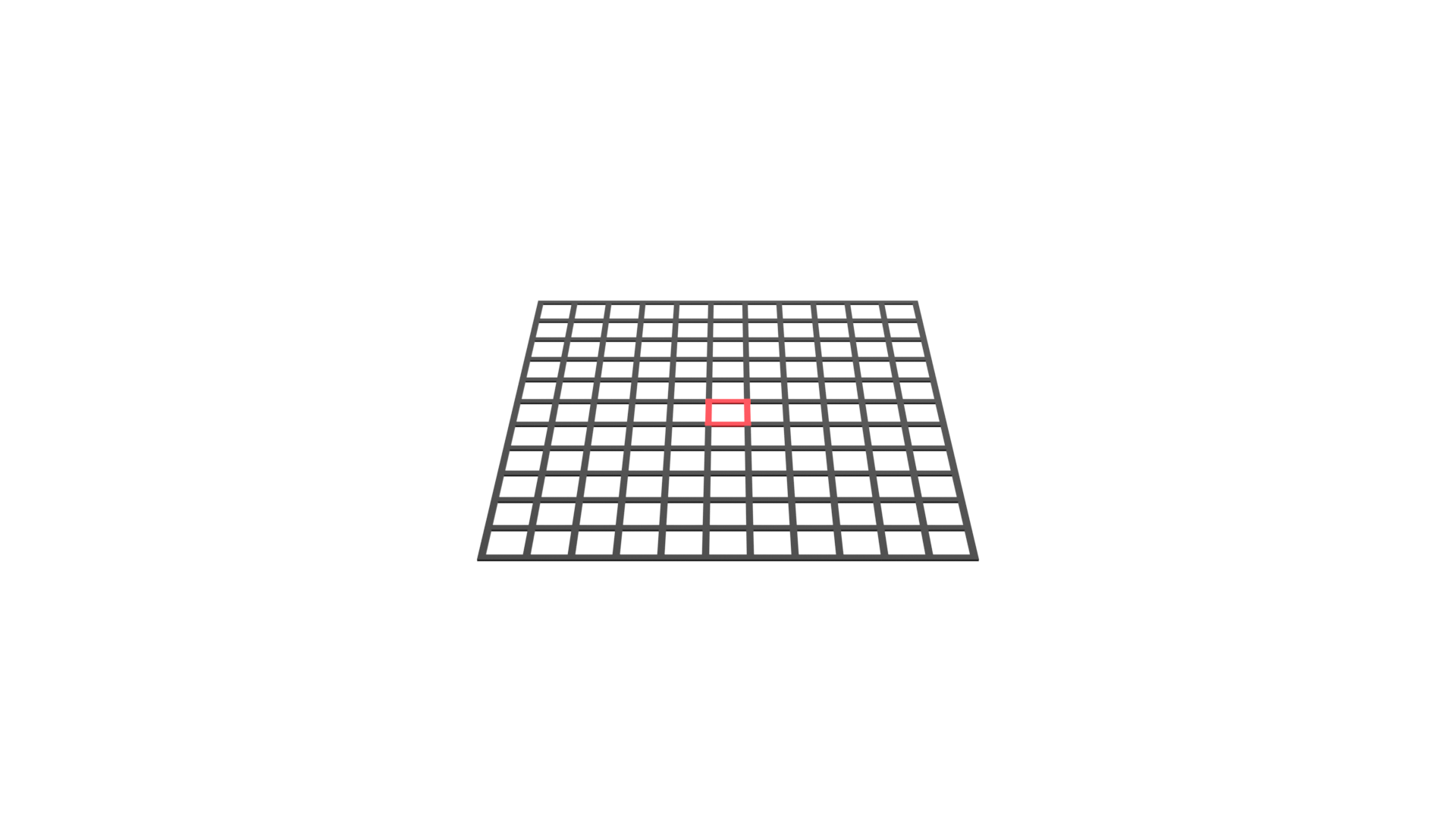}
    \motivation{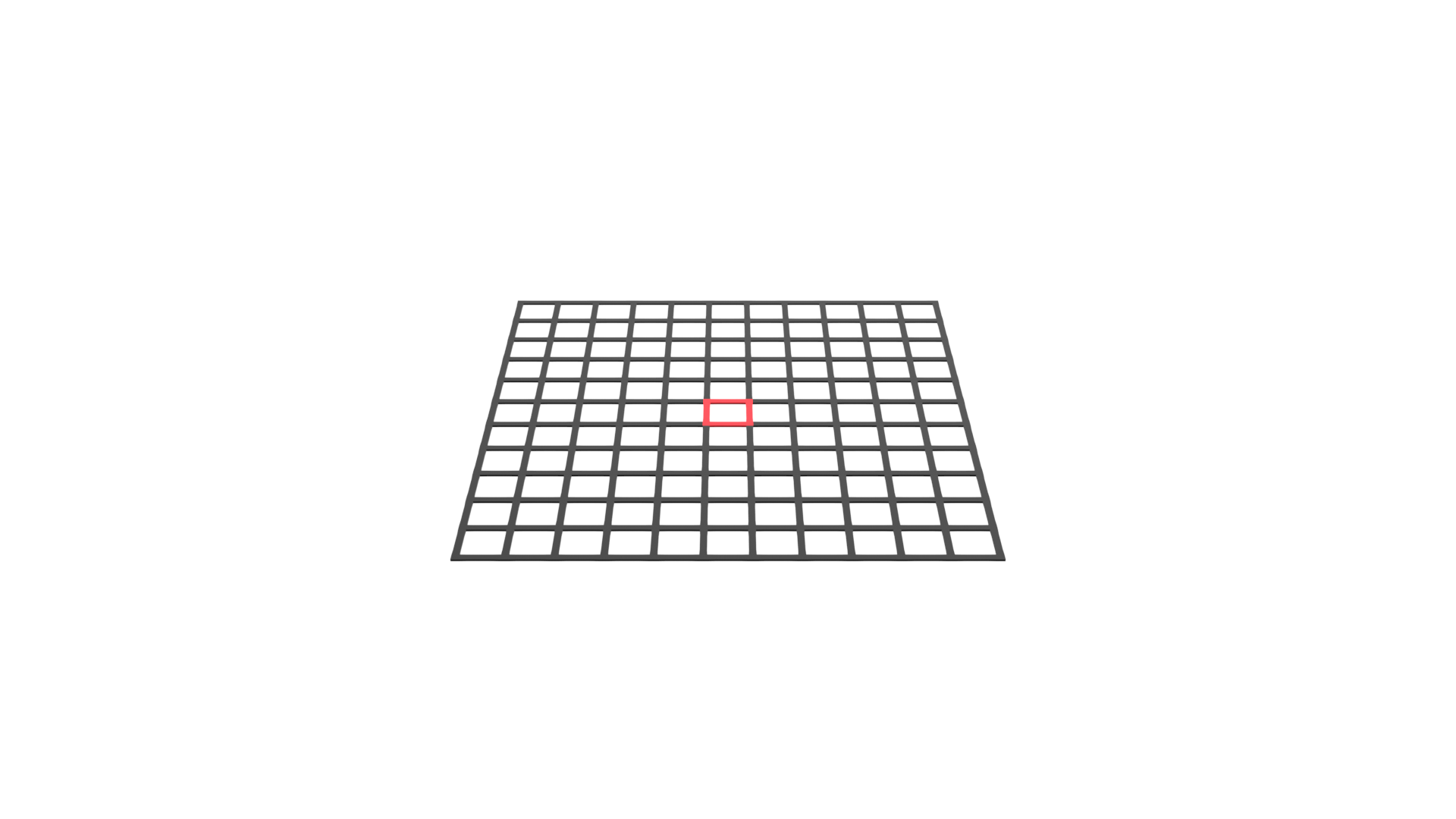}
    \motivation{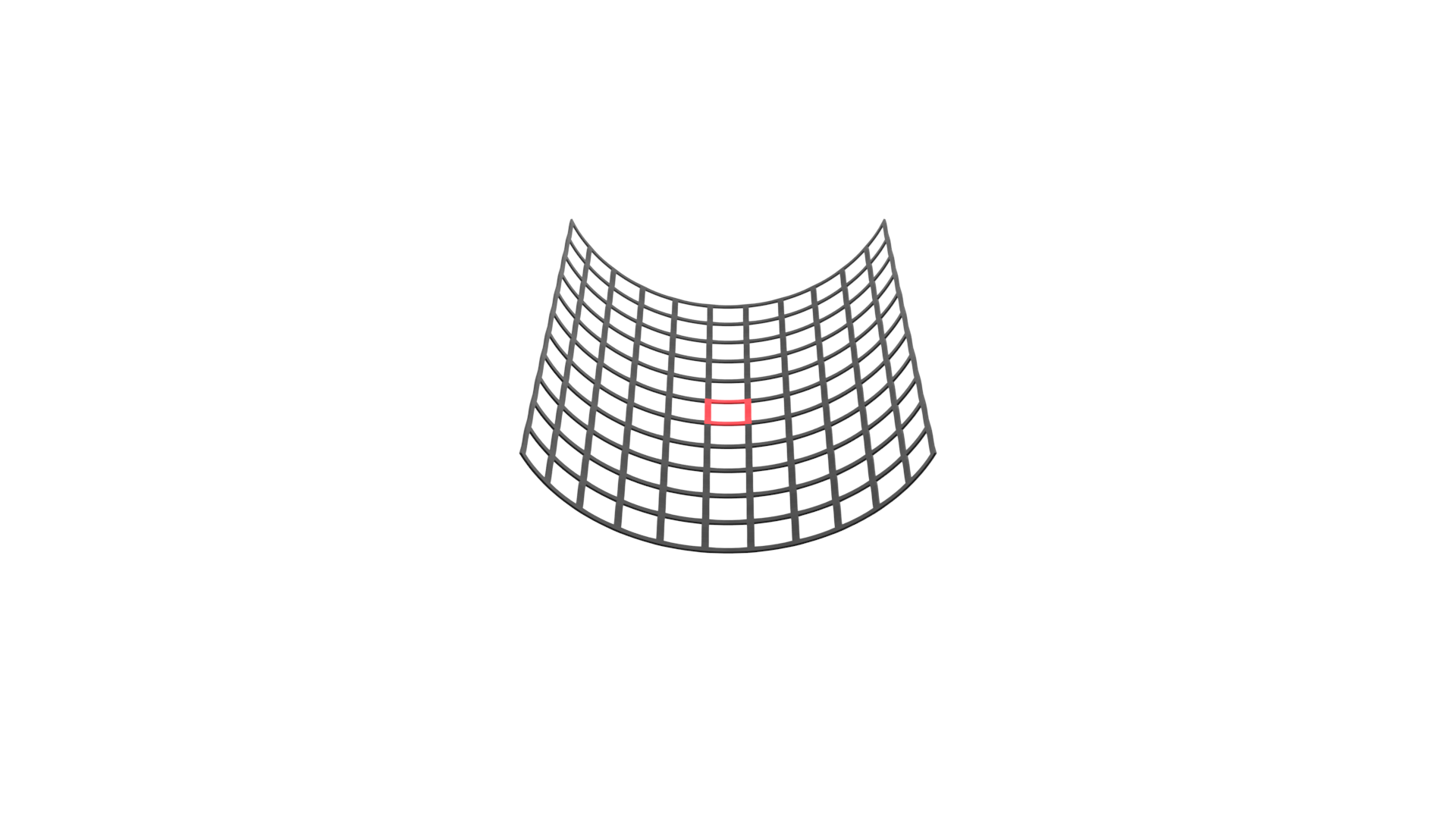}
    \caption{The same microstructure (left) is stretched $10\%$ (middle) and bend (right).
    Energy is $100\times$ higher under stretch, which indicates little correlation between energy and visual deformation.}
\Description[Energy difference in stretch vs. bending]{Image showing stretch and bending deformations.}
\label{fig:EnergyMotivation}
\end{figure}

Miguel et al.~\shortcite{Miguel2016} fitted energy functions to deformation examples.
This approach combines conservativeness with a visually meaningful fit; nevertheless, the optimization process is challenging and prone to local minima.
The authors partially circumvented this issue by formulating additive energy models in an incremental fashion.
In our numerical homogenization approach, training data can be produced under much more controlled conditions, and the optimization process is simplified.
Xu et al.~\shortcite{Xu2015}  designed conservative materials following the Valanis-Landel assumption, with energy terms dependent on principal strains.
Our eigen-strain representation also uses principal strains, but it includes principal directions to form a polar-coordinate representation for anisotropy.

In contrast to fitting energy values, fitting stress values allows one to account for the mechanical response on each deformation mode separately, hence it leads to better correlation with visual deformation.
Several authors parameterized the strain-stress response by interpolation functions, e.g., Lamé parameters~\cite{Bickel2009}, the stiffness tensor~\cite{Wang2011}, or stress values directly~\cite{Wang2020}.
Unfortunately, none of these methods is conservative.
We would like to fit stress values by parameterizing an energy function, to achieve both good correlation with visual deformation and conservativeness.
However, na\"ive fitting of stress values starting from an energy spline or RBF leaves us short of parameter degrees of freedom, i.e., with thin-shell stress being a six-dimensional derivative of energy, stress is only well defined by a six-tuple of energy values.
Chan-Lock et al.~\shortcite{ChanLock2022}
 addressed this challenge with high-order RBF interpolants, but they used it only for three-dimensional in-plane strain.
We extend high-order RBF interpolants to more complex domains, showing accurate fitting of both in-plane and out-of-plane deformation.

Neural networks have also been used for strain-dependent material parameterization.
The approach has a long history in computational mechanics~\cite{Ghaboussi1991,Shen2005,Le2015,Colasante2016,Xu2021}.
Some of these works even model conservative materials~\cite{Masi2021,Linka2021}.
The recent work of Li et al.~\shortcite{Li2023} is particularly interesting for metamaterial design, as it models the mechanical response of material families with a single neural network.
Neural networks offer an interesting approach to material parameterization due to their flexibility, but they suffer from parameter explosion and they typically lack explainability.
Our approach, based on RBF interpolants, provides closer insight into the material behavior.
One interesting approach for future work would be to use our high-order polar interpolants as building block for neural material design.

To conclude this discussion, it is worth noting that multiple works on microstructure characterization or design have used various forms of homogenization, as a way to condense their complex material response~\cite{Schumacher2018,sperl2022eylsmpf,Zhang2023,Li2023}.
Our work introduces a new material parameterization, aimed at improving the accuracy of previous homogenization methods, and can serve as a building block for material design techniques.

%% file: includes/thinshell.tex
\section{Thin-Shell Eigen-Strain}
\label{sec:thinshell}

In this section, we describe our choices of membrane and bending deformation metrics.
Both use a representation based on eigen-analysis, with principal directions of deformation, which has notable advantages for the design of interpolation-based material functions.
We start the section summarizing the choice of fundamental deformation metrics.
Then, we describe the eigen-representation, and we discuss its advantages and challenges.

\subsection{Membrane and Bending Deformation}

To describe membrane (in-plane) deformation, we start from the well-known quadratic Green strain.
With deformation gradient $F$, Green strain is defined as:
\begin{equation}
\label{eq:Green}
    E = \frac{1}{2} \, \left( F^T \, F - I \right).
\end{equation}

To describe bending (out-of-plane) deformation, we start from the triangle-averaged shape operator \cite{Grinspun:2006:CDS}:
\begin{equation}
\label{eq:ShapeOp}
\Lambda = \sum_i \frac{\xi_i}{2 \, A \, l_i} t_i \otimes t_i,
\end{equation}
with $A$ the triangle area, $\{ \xi_i \}$ edge-based signed dihedral angles, $\{ l_i \}$ edge lengths, and $\{ t_i \}$ outward-pointing edge vectors.
Given edge vectors $\{ e_i \}$ and triangle normal $n$, we compute $l_i = |e_i|$ and $t_i = e_i \times n$.
All values are measured in the deformed setting.

\subsection{Eigen-Representation of Strain}

We use an eigen-representation of strain for both membrane and bending deformations.
Let us denote either strain as $S \in \{ E, \Lambda \}$.
Given its eigen-decomposition, we denote the eigen-values as $\lambda, \mu,$ $\| \lambda \| > \| \mu \|$.
We characterize the strain $S$ as a 3-vector $(\lambda, \mu, \theta)$, where $\lambda$ and $\mu$ are, by definition, the principal strains, and $\theta$ is the angle of the eigen-vector of $\lambda$, i.e., the principal strain direction.

\begin{figure}[t!]
\centering
\includegraphics[width =\linewidth]{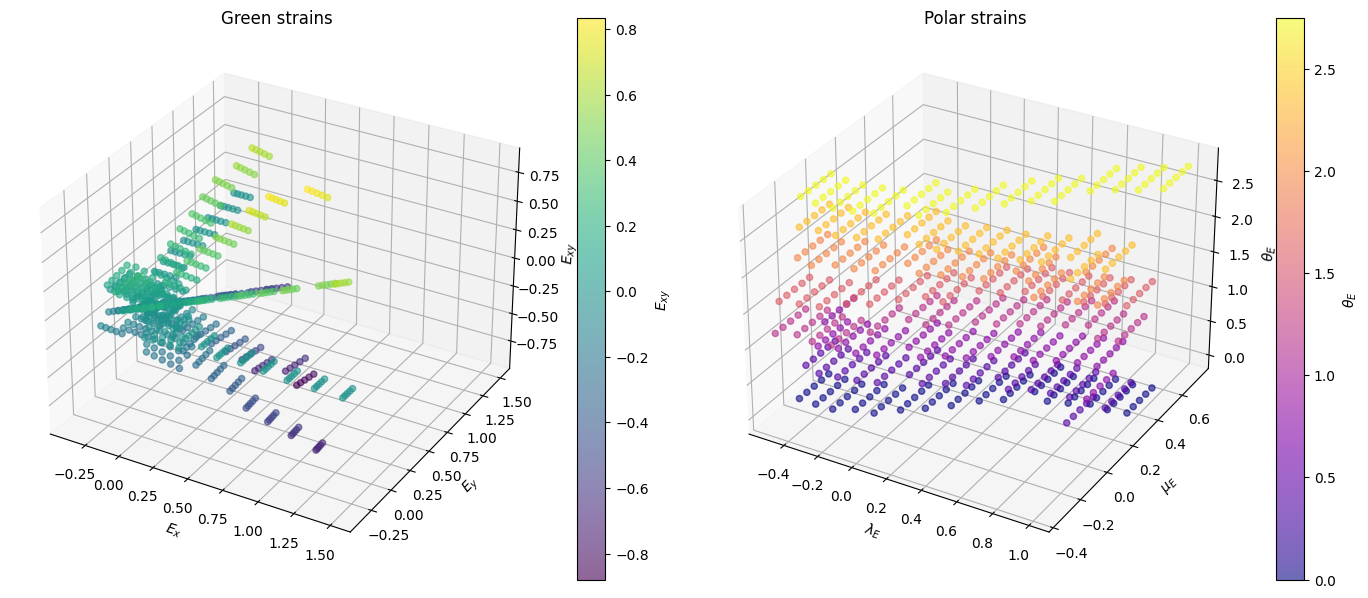}
\caption{
\label{fig:dataDistribution}
The distribution of training data is more regular and better suited for RBF interpolation on the eigen-strain domain $(\lambda_E, \mu_E, \theta_E)$ (right) than on the Green-strain domain $E$ (left).
}
\Description[Data distribution in eigen-strain vs. Green strain]{Image showing the distribution of training data in eigen-strain domain vs. Green strain domain.}
\end{figure}

There are two main advantages of using the eigen-representation of strain as the domain for interpolation-based material design.
The first advantage is that the representation is naturally well suited for capturing anisotropic material response, i.e., direction-dependent differences in material stiffness.
The second advantage is that the homogenization training data fits more compactly the strain domain.
\reffig{dataDistribution} compares the in-plane training strains of our homogenization process in eigen-domain $(\lambda_E, \mu_E, \theta_E)$ vs. regular Green-strain domain $(E_{xx}, E_{yy}, E_{xy})$.
The eigen-representation produces a more regular and less curved distribution of the training data, and this leads to higher accuracy in material estimation.

We are not the first to choose an eigen-representation of strain, but we find the treatment of this representation in prior work too simplistic.
Wang et al.~\shortcite{Wang2011} used the directions of principal stretch and principal curvature to interpolate in-plane stiffness tensors and bending stiffness separately.
However, their model is not conservative and they ignore force Jacobian terms due to changes in stiffness.
Therefore, they never cared to compute derivatives wrt the direction of strain.
Sperl et al.~\shortcite{sperl2020hylc} used the direction of principal curvature to interpolate bending energies, but they proposed this only for runtime evaluation, not during the material fitting process.
In \refsec{experiments} we demonstrate the improved accuracy of our model.

The eigen-representation of strain comes with one challenge though: the periodic nature of the principal strain direction $\theta$.
The design of interpolation-based materials cannot be addressed from a Euclidean perspective, and it requires instead special treatment of strain distances on a periodic domain.
In the next section, we present our key novel development, polar high-order interpolants, which enable stress fitting of conservative material functions on periodic domains.

%% file: includes/interpolants.tex
\section{Polar Interpolants}
\label{sec:interp}

In our design of interpolation-based thin-shell energy functions, we wish to follow the methodology of high-order RBF interpolants.
However, RBFs require special distance functions on periodic domains; therefore, it is necessary to extend the fundamental theory of high-order RBF interpolants to such periodic domains.

We start the section with a short review of high-order RBF interpolants in regular Cartesian coordinates.
Then, we derive our new polar high-order RBF interpolants.
With these ingredients, we address the design of interpolation-based thin-shell energies for membrane and bending deformation.
Our narrative combines motivation, formalization of variables and functions, and discussion.
In the supplementary document, on the other hand, we include a more implementation-oriented description of the energy model.

\subsection{Cartesian High-Order RBF Interpolants}

The development of matrix-valued RBFs~\cite{Narcowich1994,Fuselier2008} considers interpolants of the type:
\begin{equation}
\label{eq:matrixRBFinterpolant}
    \psi = w^T \, \nabla \Phi(r),
\end{equation}
where $w$ is a vector of RBF weights, $\Phi$ is an RBF, and $r = \| u \|$ with $u = \Delta x = x - x_i$ the regular difference vector for Cartesian coordinates (between the point of evaluation of the RBF $x$ and the RBF center $x_i$).
This definition of RBFs yields a vector function $v = \nabla \psi$.
The key difference with scalar RBFs is in the vector-valued weights, which provide the necessary degrees of freedom for interpolating vector fields, while starting from scalar interpolants, i.e., they are conservative by definition.

Chan-Lock et al.~\shortcite{ChanLock2022} showed that the scalar interpolant \refeq{matrixRBFinterpolant} can be recast using RBFs directly, instead of RBF gradients as:
\begin{equation}
\label{eq:cartesianRBFinterpolant}
    \psi = w^T \, u \, \phi(r).
\end{equation}
Their demonstration is based on redefining the RBF $\phi(r) = \frac{\Phi'(r)}{r}$.
They also extended the methodology beyond vector-field interpolation, and they defined high-order interpolants for higher-order derivatives of conservative fields (e.g., Hessians).
We do not use this aspect of their work, as we found that fitting Hessians did not increase accuracy with respect to the parameter count, probably because we fit energies in 5D vs. their 3D work.

\subsection{Polar High-Order RBF Interpolants}

We aim to extend the method of high-order RBF interpolants to polar coordinates, i.e., coordinate sets that combine linear coordinates and periodic angular coordinates.
We start by defining a new difference vector, which we refer to as {\em polar difference vector}:
\begin{equation}
    u = \begin{cases}
    \Delta x, & \text{ if } x \text{ is linear.} \\
    \sin(\Delta x), & \text{ if } x \text{ is angular.}
    \end{cases}
\end{equation}
For angular coordinates, we use a difference function that is smooth and evaluates to 0 at 0 and $\pi$.
This ensures continuity and smoothness of the difference function for periodic angular coordinates.

Starting from the regular matrix-valued RBF interpolant \refeq{matrixRBFinterpolant}, we evaluate the RBF gradient with our new polar difference vector:
\begin{equation}
\label{eq:matrixRBFinterpolantNew}
    \psi = w^T \, \Phi'(r) \, \nabla u \, \frac{u}{r}.
\end{equation}

Same as Chan-Lock et al., we redefine the RBF $\phi(r) = \frac{\Phi'(r)}{r}$.
Then, we obtain the following polar high-order RBF interpolant:
\begin{align}
\label{eq:polarRBFinterpolant}
    & \psi = w^T \, \nabla u \, u \, \phi(r), \\
    \nonumber
    & \text{with } \text{diag}(\nabla u) = \begin{cases}
    1, & \text{ if } x \text{ is linear.} \\
    \cos(\Delta x), & \text{ if } x \text{ is angular.}
    \end{cases}
\end{align}

The resulting polar high-order interpolant resembles strongly the Cartesian high-order interpolant \refeq{cartesianRBFinterpolant}, with two key differences for angular coordinates: the new difference function $\sin(\Delta x)$, and a weight $\cos(\Delta x)$.
These subtle but key changes allow us to use the power of high-order RBF interpolants together with the material design accuracy of eigen-representations of strain.



\subsection{RBF Energy Model}
\label{sec:interp:RBF}

The high-order polar RBF interpolants are the main ingredient to design a general interpolation-based energy for thin-shell membrane and bending deformation.
Nevertheless, we need to address two additional questions.
First, we need to define the specific interpolation domain for the fitted energy function, and we do this with awareness of the available training data.
Second, we need to define the energy behavior under isotropic strain ($\lambda = \mu$), where the eigen-strain is not differentiable because $\theta$ is undefined, and we do this by blending anisotropic and isotropic energy evaluations.

In defining the eigen-strain domain and the full energy model, we follow this notation.
We denote as $\hat{x} = (\lambda_E, \mu_E, \theta_E, \lambda_\Lambda, \mu_\Lambda, \theta_\Lambda)$ $\in \R^6$ the full thin-shell eigen-strain, as described in \refsec{thinshell}.
However, as emphasized by Sperl et al.~\shortcite{sperl2020hylc}, the training data for homogenization is limited to a 5D subspace with $\mu_\Lambda = 0$ (i.e., uniaxial/cylindrical bending, see more details in \refsec{homogenization:data}).
Then, we define a cylindrical-bending eigen-strain domain $\bar{x} = (\lambda_E, \mu_E, \theta_E, \lambda_\Lambda^*, \theta_\Lambda^*)$ $\in \R^5$.
Details on the values of the bending strain $(\lambda_\Lambda^*, \theta_\Lambda^*)$ will follow.
Finally, we normalize the 
principal strain components $\lambda_E, \mu_E, \lambda^*_\Lambda$ based on the range of the training data, to ensure that all training strains are in the range $\left[ -1, 1 \right]$.
The polar difference operator achieves this for angular strains, hence no normalization is necessary.
As a result, we obtain the RBF interpolation domain $x \in \R^5$.
In this normalized strain domain, we construct fundamental RBF interpolants $\{ \psi_i \}$ following \refeq{polarRBFinterpolant}, and we combine them to produce the fitted energy model which we denote as $\Psi$.
Finally, we define a full energy model $\Psi_\text{full}$ to robustly support isotropic deformations, which details described next.

We follow the approach of Sperl et al.~\shortcite{sperl2020hylc} to address the dimmnsionality mismatch of 5D training data $\bar{x}$ vs. 6D runtime strain $\hat{x}$.
We evaluate each RBF interpolant $\psi_i$ (with the same RBF center and weights) for both principal curvatures.
This yields the fitted energy model:
\begin{align}
\label{eq:fundamentalEnergy}
    \Psi = \sum_i & \psi_i(\lambda_E, \mu_E, \theta_E, \lambda_\Lambda^* \leftarrow \lambda_\Lambda, \theta_\Lambda^* \leftarrow \theta_\Lambda) \\
    \nonumber
    + & \psi_i(\lambda_E, \mu_E, \theta_E, \lambda_\Lambda^* \leftarrow \mu_\Lambda, \theta_\Lambda^* \leftarrow \theta_\Lambda + \frac{\pi}{2}).
\end{align}

Previous interpolation-based energy models~\cite{Miguel2016, sperl2020hylc} typically defined interpolants in low-dimensional subspaces, and then combined multiple energy terms in an 
additive manner.
We find that this is not necessary for our homogenization methodology, and we can design a 6D energy directly, thanks to our polar interpolation approach.

The fitted energy \refeq{fundamentalEnergy} is robust under anisotropic strain ($\lambda \neq \mu$).
However, as mentioned above, it is not differentiable in the isotropic case ($\lambda = \mu$), where $\theta$ is undefined.
For runtime evaluation, we handle this through localized blending of isotropic evaluations of the fitted energy~\refeq{fundamentalEnergy}, forcing the principal strain direction $\theta = 0$.
Then, our full thin-shell energy model becomes:
\begin{align}
\label{eq:completeEnergy}
\nonumber
    \Psi_\text{full} &= (1 - \omega(\lambda_E, \mu_E)) \, (1 - \omega(\lambda_\Lambda, \mu_\Lambda)) \, \Psi(\lambda_E, \mu_E, \theta_E, \lambda_\Lambda, \mu_\Lambda, \theta_\Lambda) \\
\nonumber
    &+ \omega(\lambda_E, \mu_E) \, (1 - \omega(\lambda_\Lambda, \mu_\Lambda)) \, \Psi(\lambda_E = \mu_E, \theta_E = 0, \lambda_\Lambda, \mu_\Lambda, \theta_\Lambda) \\
\nonumber
    &+ (1 - \omega(\lambda_E, \mu_E)) \, \omega(\lambda_\Lambda, \mu_\Lambda) \, \Psi(\lambda_E, \mu_E, \theta_E, \lambda_\Lambda = \mu_\Lambda, \theta_\Lambda = 0) \\
    &+ \omega(\lambda_E, \mu_E) \, \omega(\lambda_\Lambda, \mu_\Lambda) \, \Psi(\lambda_E = \mu_E, \theta_E = 0, \lambda_\Lambda = \mu_\Lambda, \theta_\Lambda = 0),
\end{align}
with a weight function $\omega(\lambda, \mu) = \text{e}^{-(\lambda - \mu)^\gamma}$ that localizes the isotropic evaluations.
It is not necessary to design multiple energy terms, we fit just a single energy~\refeq{fundamentalEnergy}, and then we blend four evaluations of this energy at runtime: anisotropic stretch and bending, isotropic stretch, isotropic bending, and isotropic stretch and bending.

The final energy model $\Psi_\text{full}$ does not guarantee zero stress at zero strain by construction.
We could add the zero-strain stress compensation term of Chan-Lock et al.~\shortcite{ChanLock2022} a posteriori of model fitting, but we found this to be unnecessary in practice, because the fitted stress at zero strain was negligible in all our experiments.

%% file: includes/homogenization.tex
\section{Homogenization Workflow}
\label{sec:homogenization}

In this section, we describe the data generation and fitting process.
Our approach to data generation is largely similar to previous work based on simulation with periodic boundary conditions, which simplifies the exploration of continuum strain values.
In the description of the energy-model fitting process, we discuss data normalization and the optimization methodology, but we refer to the supplementary document for full details.

\subsection{Generation of Training Data}
\label{sec:homogenization:data}

To generate training data, we follow a controlled simulation approach.
In a nutshell, we simulate microstructure deformations under homogeneous coarse strain $x$ using periodic boundary conditions, and we compute the homogenized stress $\DD{\Psi}{x}$.
In this way, we obtain data pairs $(x, \DD{\Psi}{x})$, which we use for fitting our RBF energy model, as we discuss in detail in the next subsection.

We sample the strain domain $x$ in the following way.
For stretch, we start by sampling the first principal stretch $\lambda_E$ and the principal direction $\theta_E$.
For each pair, we obtain the minimum-energy orthogonal stretch, and we sample the second principal stretch $\mu_E$ in a positive and negative range around this minimum-energy configuration.
For bending, as discussed by Sperl et al.~\shortcite{sperl2020hylc}, it is not possible to apply arbitrary uniform coarse bending.
We follow their approach, and apply only uniaxial bending, i.e., we sample the first principal curvature $\lambda_\Lambda$ and the principal direction $\theta_\Lambda$, with the other principal curvature $\mu_\Lambda = 0$.
To summarize, for each microstructure, we sample regularly a 4D domain $(\lambda_E, \theta_E, \lambda_\Lambda, \theta_\Lambda)$.
Then, for each 4D sample we adapt the sampling of $\mu_E$ around the minimum energy configuration, and we leave $\mu_\Lambda = 0$.

\begin{figure}[t!]
\centering
\includegraphics[width =.75\linewidth]{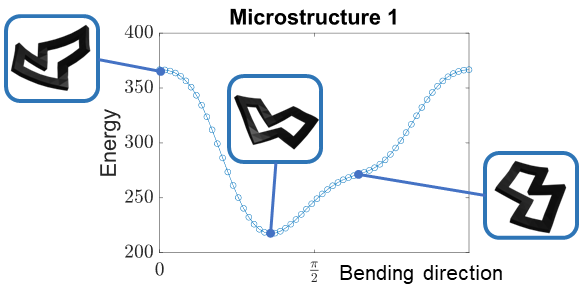}
\caption{
\label{fig:energyprofiles}
The plot shows, for Microstructure 1, the energy density wrt bending direction under uniform unaxial curvature.
This microstructure reveals a clear 90-degree asymmetry, underlining  the importance of fitting the material model on a range $\left[ 0, \pi \right)$ of principal directions.
}
\Description[Plot showing energy asymmetry of Microstructure 1]{Plot showing energy vs. bending direction for Microstructure 1, highlighting 90-degree asymmetry.}
\end{figure}

There are a few important considerations about the sampling range.
For in-plane strain, we have sampled $50\%$ compression to $100\%$ stretch, which is well beyond the range that the material can sustain in reality without breaking.
For bending strain, we define the maximum $\lambda_\Lambda$ as the value that makes the simulation tile a perfect cylinder.
As all our microstructures are homogeneous in the thickness direction, deformations are identical for positive and negative curvatures, and we sample only positive ones.
For the principal stretch and bending directions, we sample the range $[0, \pi)$.
Several authors have assumed 90-degree symmetry~\cite{Wang2011,sperl2020hylc}, but we find that this is not the case for all microstructures. 
\reffig{energyprofiles} shows the clear 90-degree asymmetry of Microstructure 1.

For the simulation of microstructures with uniform coarse deformation under periodic boundary conditions, we refer to the work of Sperl et al.~\shortcite{sperl2020hylc,sperl2022eylsmpf}.
We use a simulation domain formed by exactly one tile of the microstructure pattern, and we solve the microstructure simulation with a 3D FEM discretization and a nearly incompressible Neo-Hookean hyperelastic material model~\cite{HHAM}.
To compute the coarse stress for each coarse strain, it suffices to measure microstructure forces $f_k$ on nodes $X_k$ of the boundary of the simulation tile~\cite{Schumacher2018,sperl2022eylsmpf}: $\DD{\Psi}{x} = \sum_k f_k^T \, \DD{X_k}{x}$.

\subsection{Fitting of the Energy Model}
\label{sec:Homogenization:Fitting}

Our energy model is parameterized by the number of RBF kernels, their center locations in eigen-strain domain $\{ x_i \}$, the RBF radii, and the vector-valued weights $\{ w_i \}$.
In all our experiments, we have used Gaussian RBFs.
We approach the optimization of the energy model by staggering steps of kernel placement, radius refinement, and weight estimation.
Prior to this, we normalize all the stress data.
Recall from \refsec{interp:RBF} that the eigen-strain $x$ is also normalized based on the range of the training data.

We form a training batch with all the training data that shares the principal stretch value $\lambda_E$.
Then, within each batch, we normalize each stress component to the range $\left[ -1, 1 \right]$, based on the range of the stress component within the batch.
We form normalization batches based on $\lambda_E$ because it is the dominant strain, and otherwise the samples with low $\lambda_E$ (e.g., only bending but no stretch) suffer high relative stress error.

For fitting, we do not need the full energy model~\refeq{completeEnergy} with isotropic and anisotropic terms; it suffices to fit the fundamental energy model~\refeq{fundamentalEnergy}.
This is possible for two reasons.
One is that, in contrast to runtime simulation, there is no discontinuity issue in the principal directions.
The second one is that we augment the isotropic data $\lambda = \mu$, inputting the same stress values for multiple data entries with different strain directions $\theta$ (we use $10$ directions in practice).

Given all normalized training data pairs $(x, \DD{\Psi}{x})$, we finally fit the energy model~\refeq{fundamentalEnergy}.
For each optimization iteration, we add an RBF center based on furthest point sampling, we optimize a common RBF radius (using COBYLA), and we fit the RBF weights.
Note that we fit optimal RBF weights for each candidate RBF radius within the execution of COBYLA.
We compute these weights by minimizing the $L^2$ stress error at all the training samples, which amounts to a simple convex quadratic optimization problem on the RBF weights.
We conclude the optimization when the RMSE of stress is below $10\%$, or if the optimization process stagnates.
For our test materials, only 3 of them failed to reach the $10\%$ error threshold.
Please see a more detailed discussion in the next section.

%% file: includes/experiments.tex
\newcommand{\microtile}[1]{
\includegraphics[width =0.068\linewidth]{images/MicrostructureTiles/mat#1.png}
}

\begin{table*}[t!]
\caption{The top row shows the 10 microstructures tested in our experiments, and the next row the id corresponding to each microstructure.
We have fitted our homogenized material model to each microstructure, until the stress RMSE is below $10\%$ (or if the process stagnates).
The third row shows the number of RBFs of the resulting material model, and the fourth row the fitting error.
Finally, the fifth row shows the number of nodes of the microstructure-level simulation meshes used for controlled validation experiments.}
\centering
\begin{tabular}{|c|cccccccccc|}
\hline
& \microtile{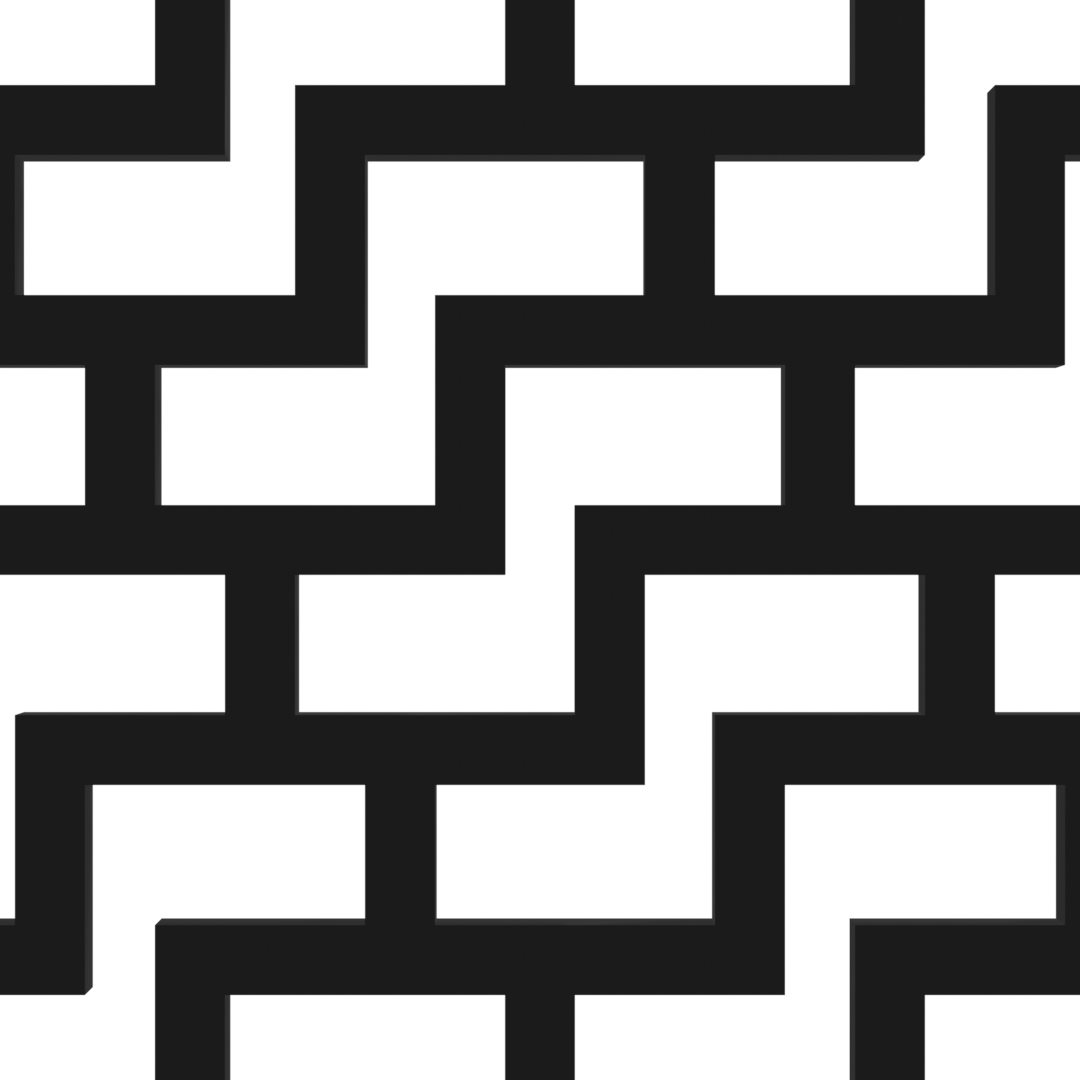}
& \microtile{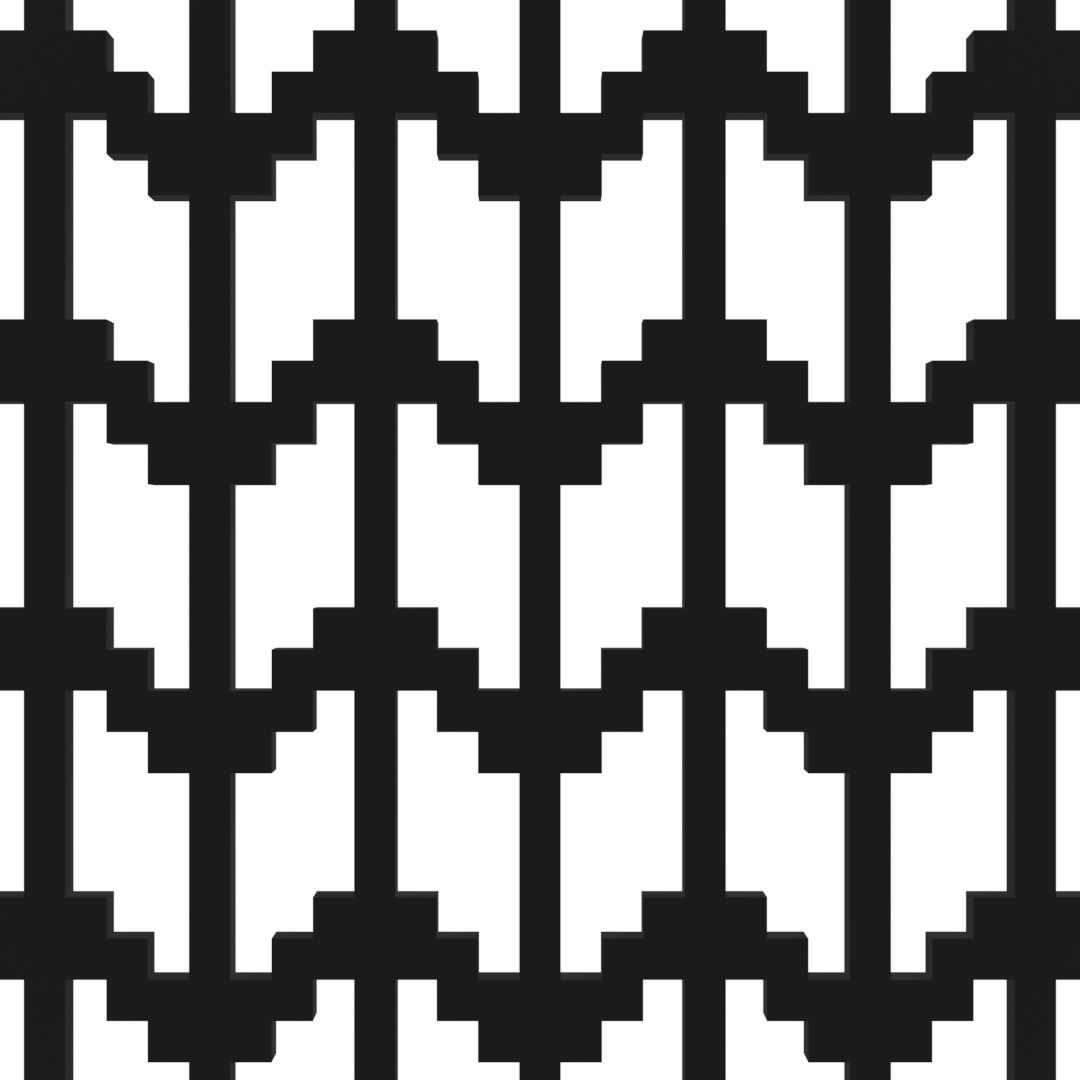}
& \microtile{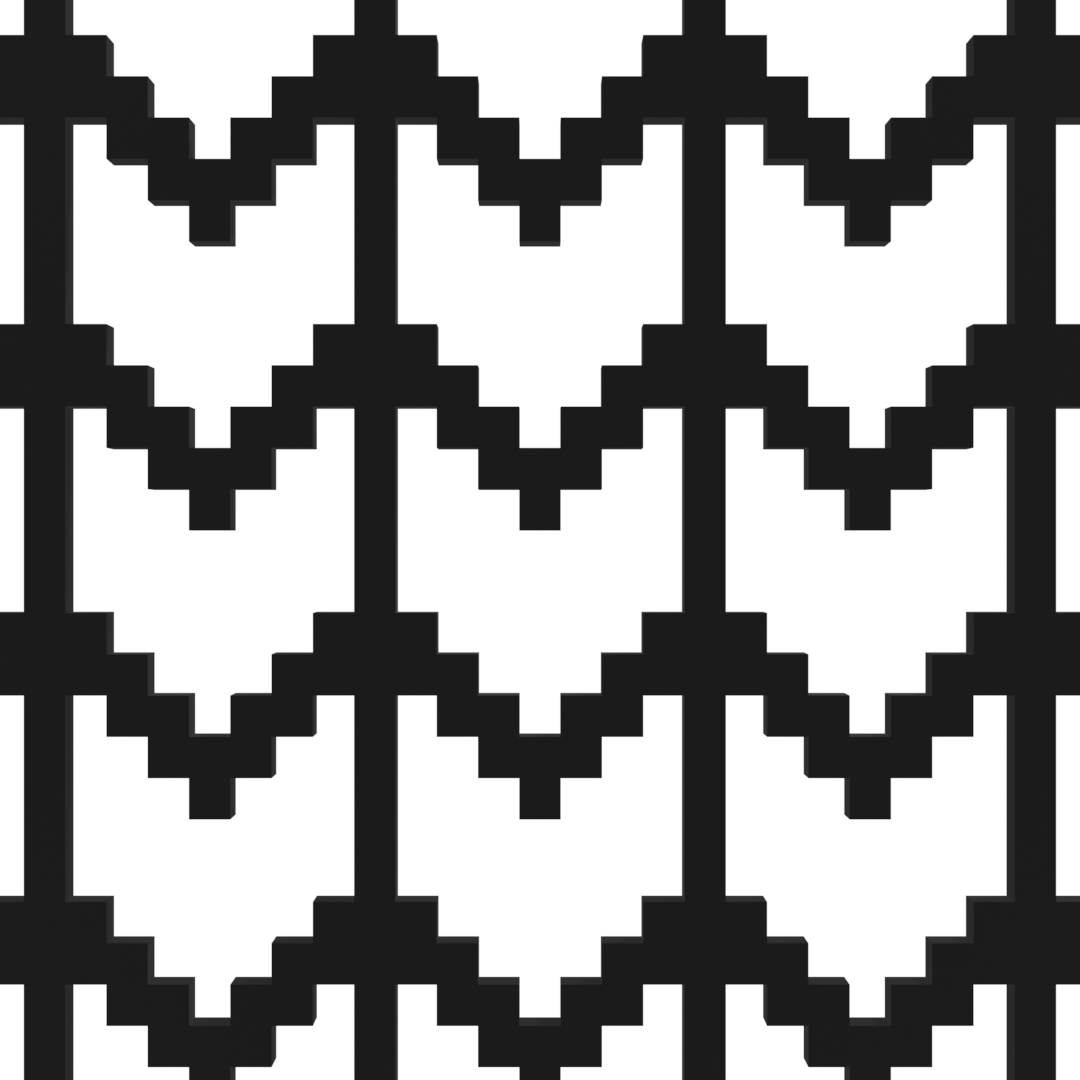}
& \microtile{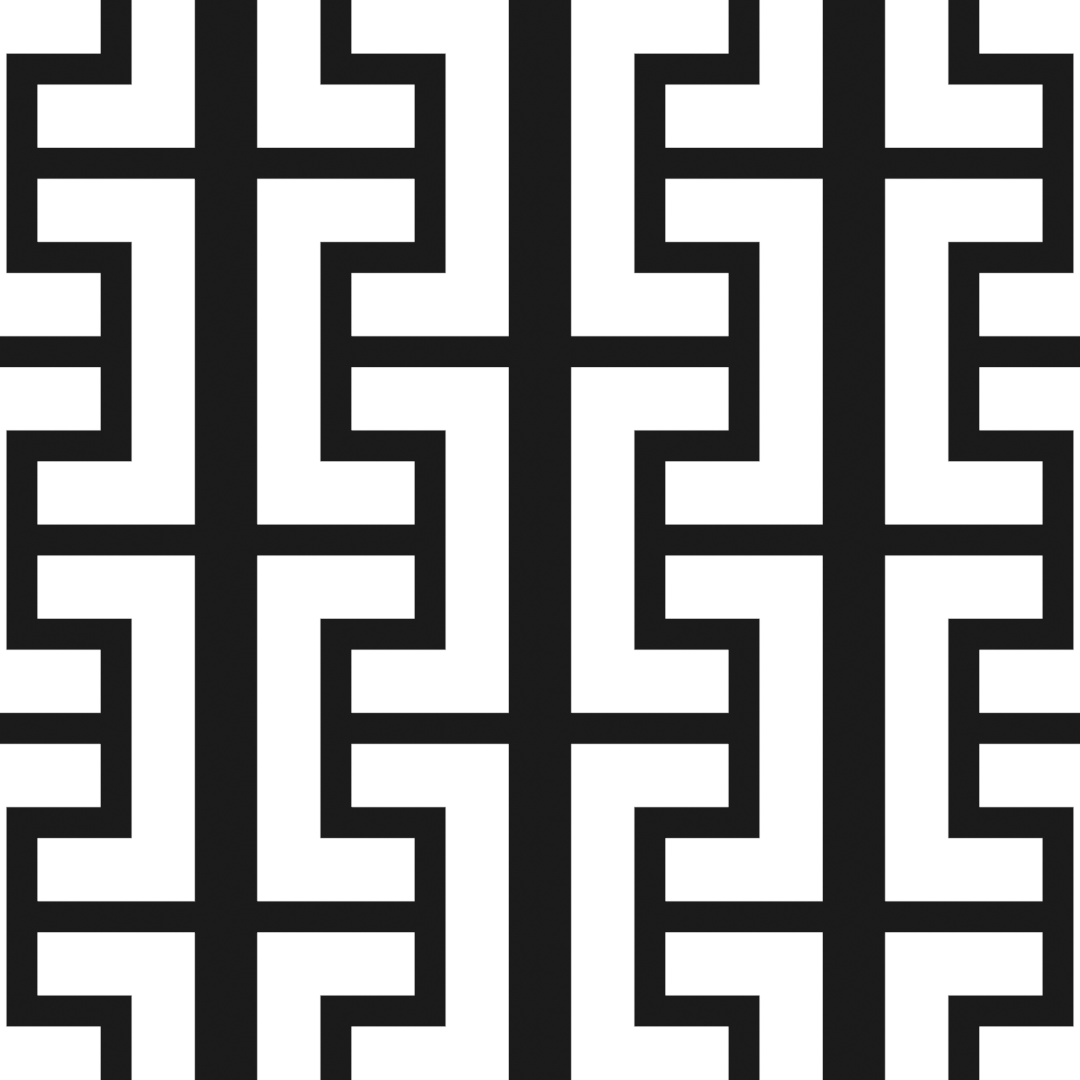}
& \microtile{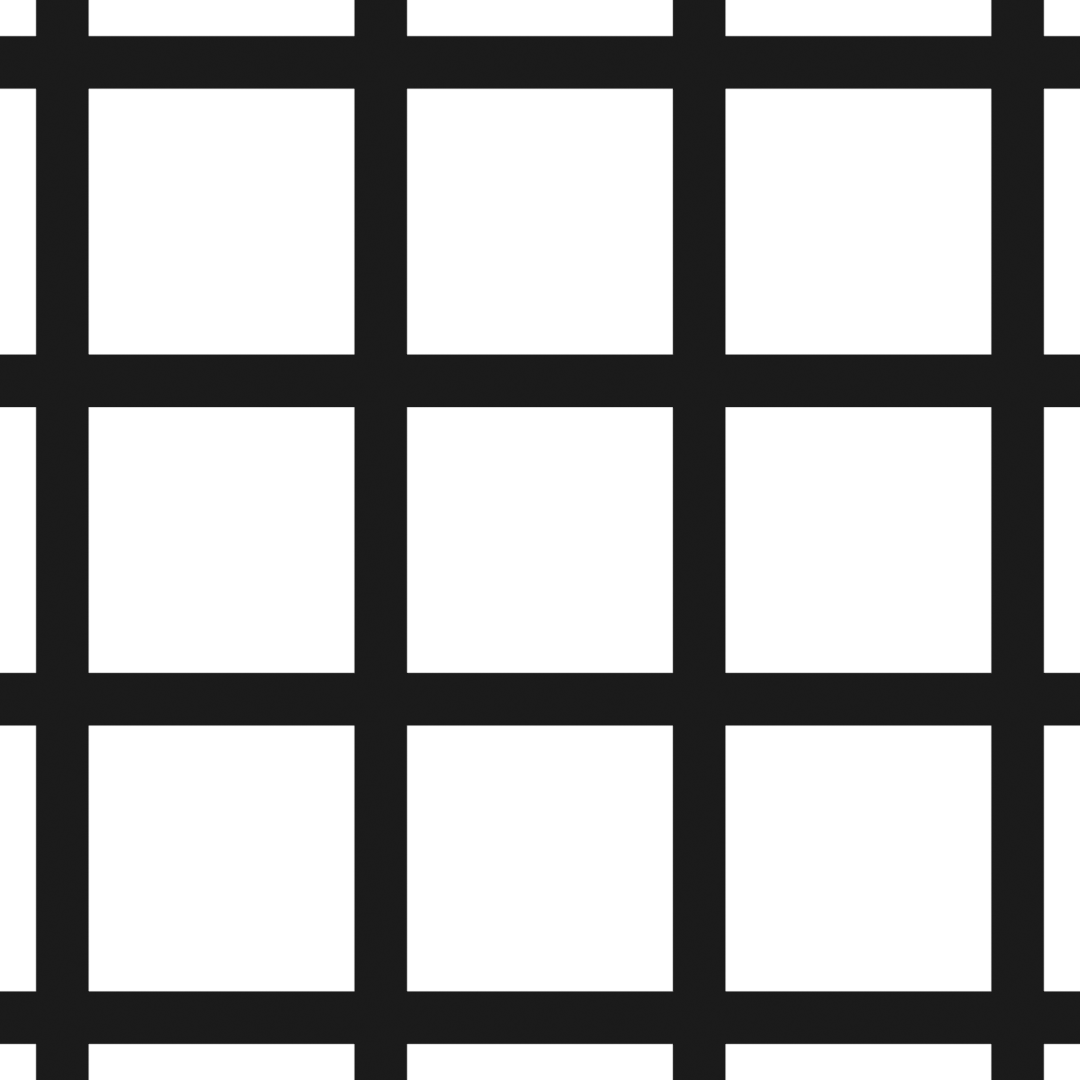}
& \microtile{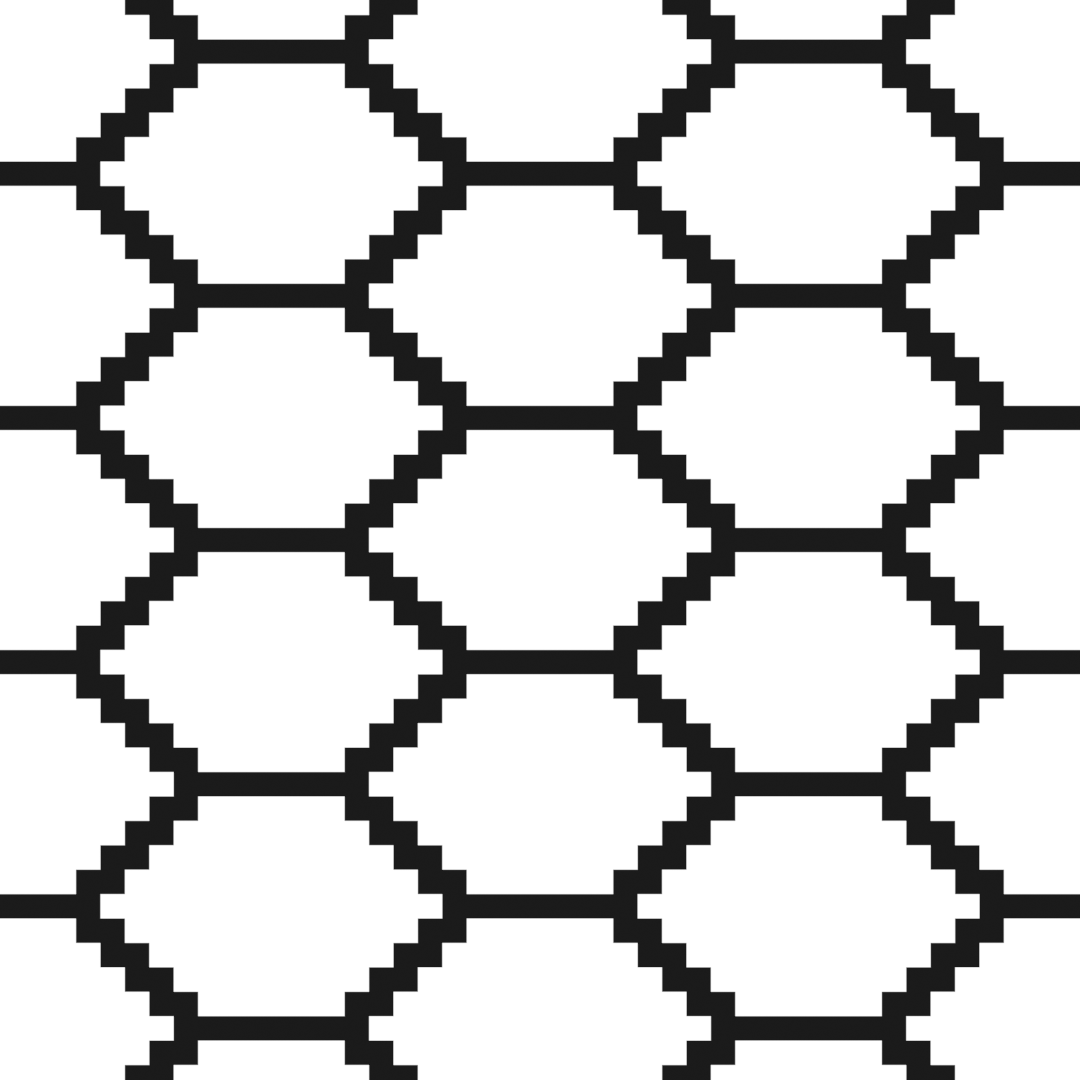}
& \microtile{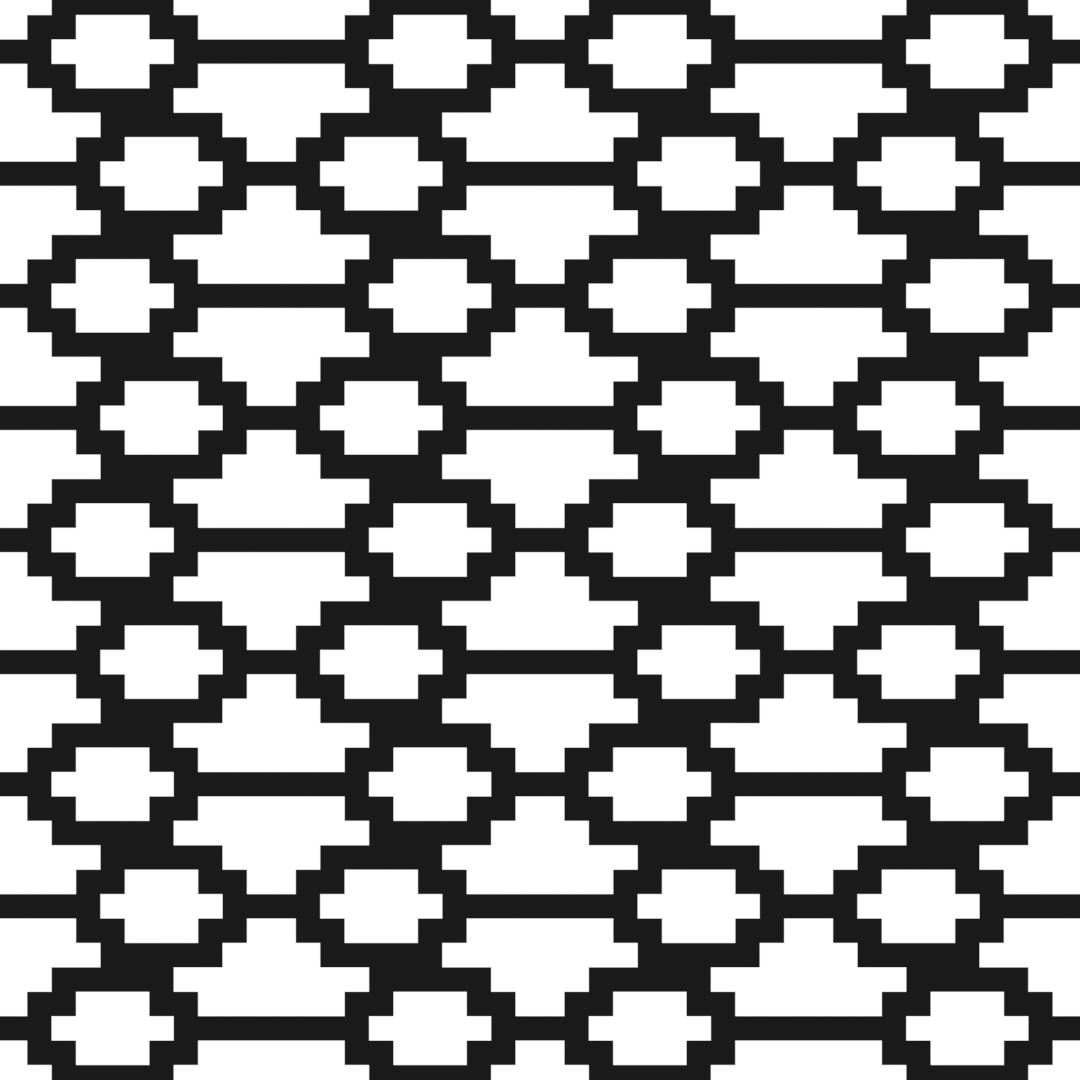}
& \microtile{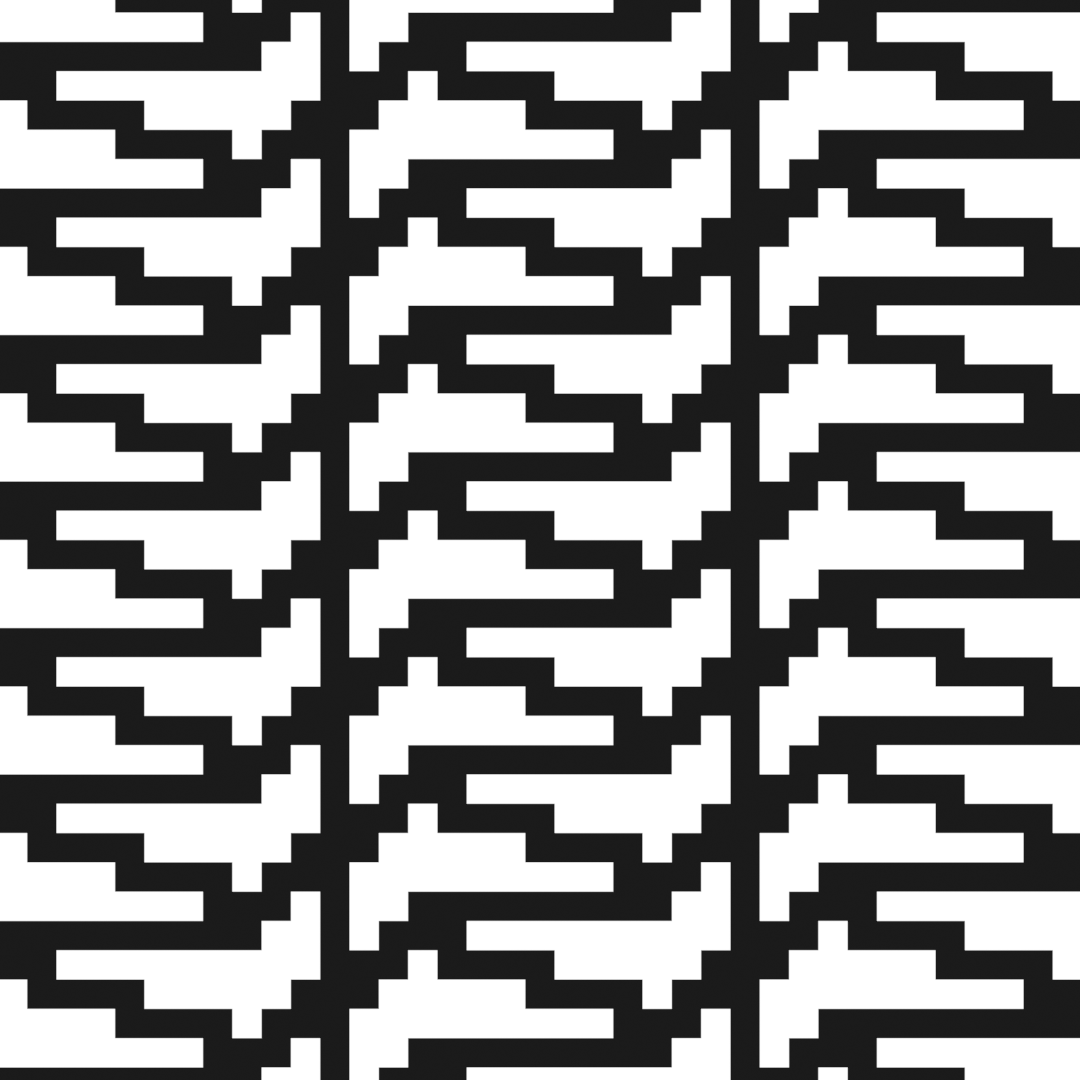}
& \microtile{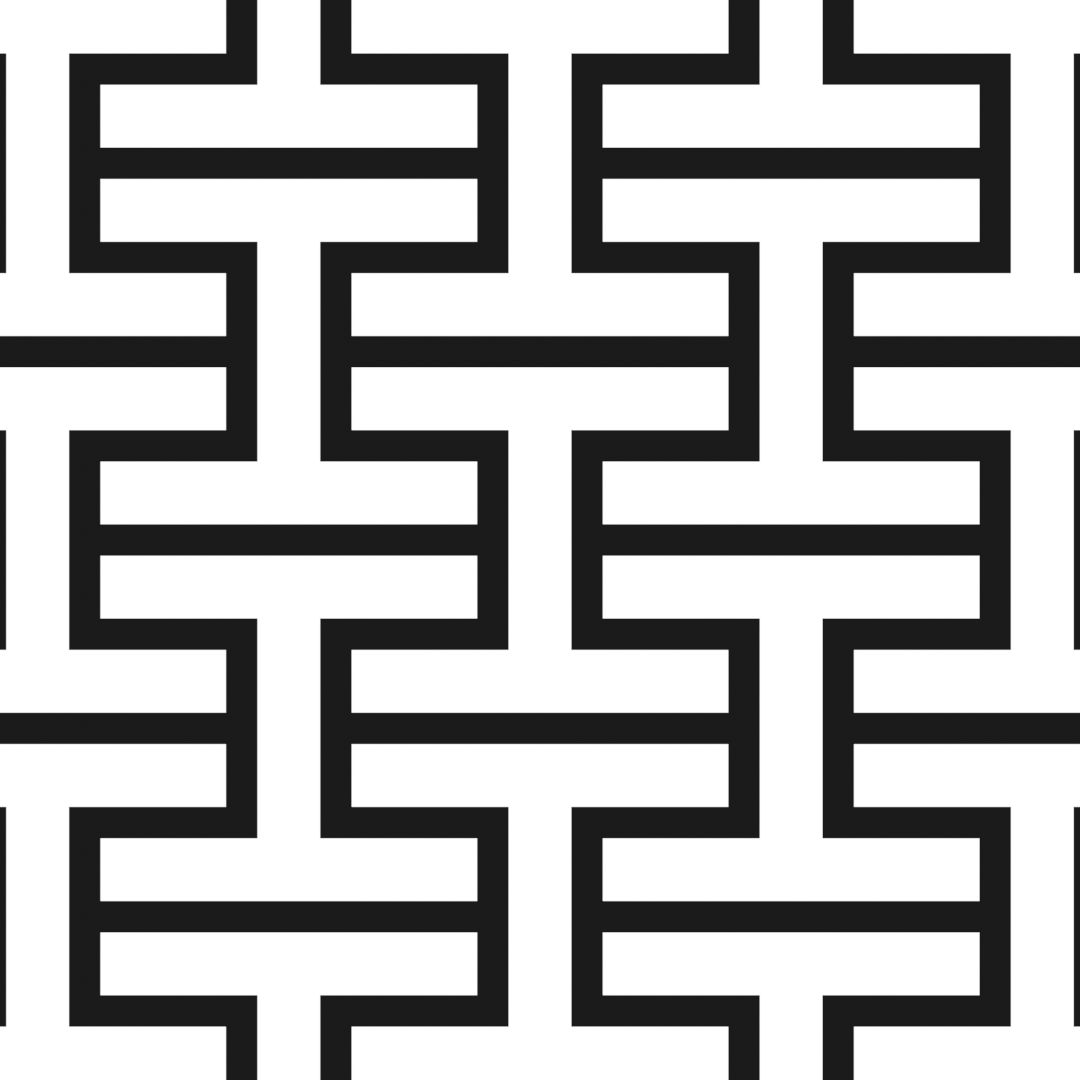}
& \microtile{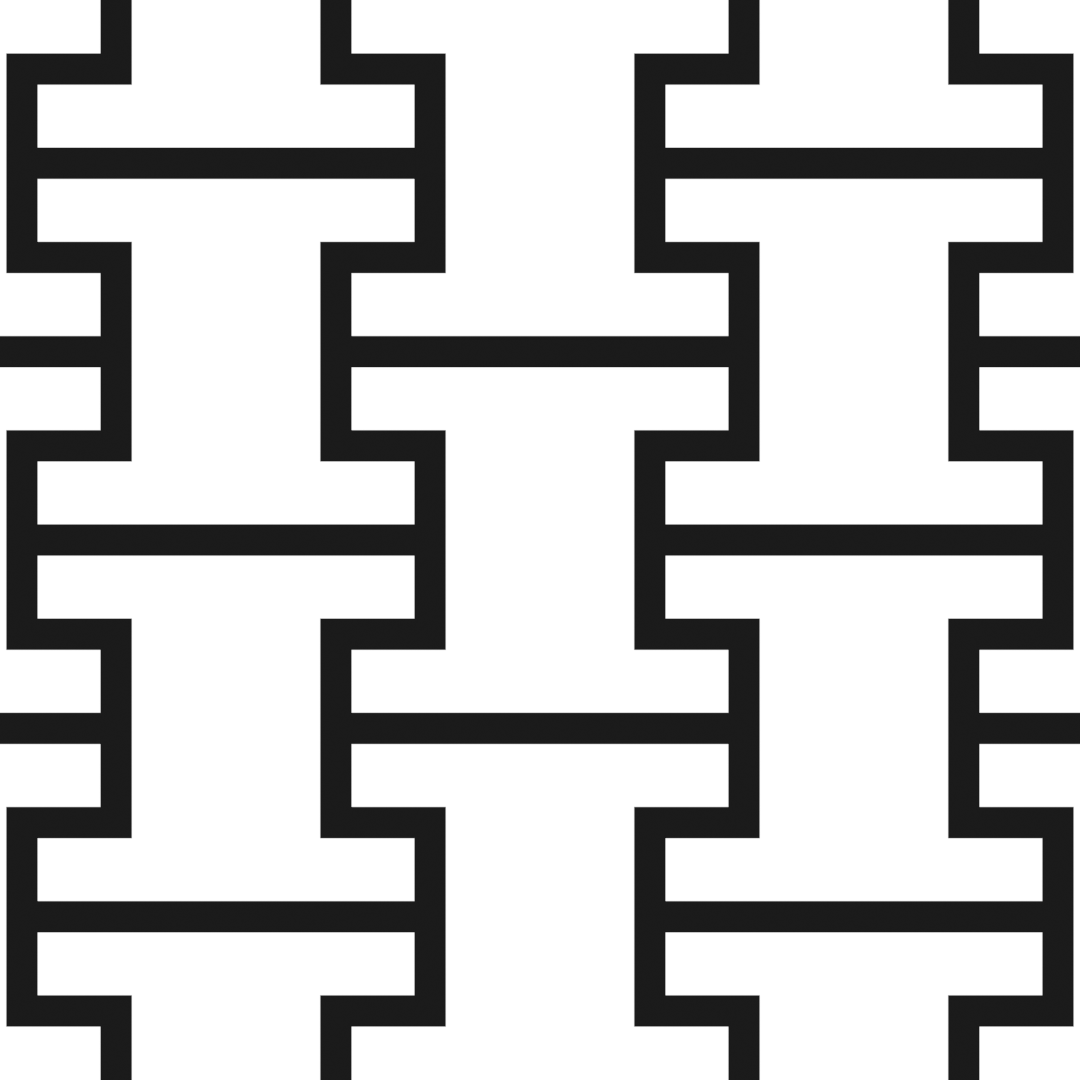} \\
\hline
Id & 1 & 2 & 3 & 4 & 5 & 6 & 7 & 8 & 9 & 10 \\
\hline
RBFs & 130 & 180 & 130 & 500 & 130 & 480 & 280 & 450 & 480 & 450 \\
\hline
Error $\%$ & 9.82 & 9.94 & 9.83 & 14.84 & 9.75 & 9.91 & 9.81 & 9.90 & 15.96 & 12.11 \\
\hline
Nodes & 17k & 27k & 22k & 34k & 6k  & 28k & 57k & 44k & 27k & 20k \\
\hline
\end{tabular}
\label{tab:Fitting}
\end{table*}

\section{Experiments and Discussion}
\label{sec:experiments}

\subsection{Implementation Details}

\reftab{Fitting} shows all 10 microstructures tested in our experiments.
For all simulations, we considered a TangoGray material.
We also fabricated 4 of the microstructures (ids: 1, 3, 4 and 5) using the TangoGray material, on an Object30 Prime Stratasys 3D printer.
As mentioned in \refsec{homogenization:data}, we have modeled the micro-structure material using a nearly-incompressible hyperelastic Neo-Hookean.
We adopt the energy density function and the material parameters for TangoGray from~\cite{HHAM}, in particular $\Psi = \frac{1}{2} \, G_0 \, (J^{-2/3} \, \text{tr}(F^T \, F) - 3) + \frac{1}{2} \, K_0 \, (J-1)^2$, $G_0 = 1.7$~MPa and $K_0 = 84.43$~MPa.
We have executed the training microstructure simulations using regular hexahedral meshes.

There are a few technical details worth noting.
For high-quality homogenized results, we have used regular equilateral meshes, which guarantee convergence of the shape operator~\refeq{ShapeOp}, as discussed in~\cite{Grinspun:2006:CDS}.
To obtain robustly the principal direction from 2D eigenvectors, we follow~\cite{486688}.
All our simulations are the result of solving for static equilibrium, and we do this with standard Newton with line search.
In the supplementary document we provide gradients and Hessians of the runtime energy model wrt eigen-strain.
For the derivative of eigen-strain wrt degrees of freedom, we have used symbolic differentiation, which can be found in our code.
As a precaution, to prevent Newton steps from reaching outside the material's domain, we add a stiff quadratic barrier energy per strain component at the maximum training strain.
Note that no simulation actually hit the boundary at equilibrium, as our training domain is conservative.

\subsection{Ablation Study}

\begin{table}[b!]
\caption{Ablation study of our model (RBF $\theta_\Lambda$, stress fit), evaluating both energy and stress error.
All fits are computed with the same number of parameters (300).}
\centering
\begin{tabular}{|c|c|c|c|c|}
\hline
& \multicolumn{2}{c|}{Linear $\theta_\Lambda$} & \multicolumn{2}{c|}{RBF $\theta_\Lambda$} \\ 
& Energy Fit & Stress Fit & Energy Fit & Stress Fit \\ \hline
Energy error & 9.19\% & 11.28\% & 4.11\% & 8.61\% \\
Stress error & 95.21\% & 31.04\% & 35.04\% & 25.17\% \\
\hline
\end{tabular}
\label{tab:EnergyFitAblation}
\end{table}

We have evaluated fitting accuracy under different model settings.
In particular, we have tested two settings used by Sperl et al.~\shortcite{sperl2020hylc}: energy fitting and linear interpolation of energy models between $0^\circ$ and $90^\circ$ in the bending direction $\theta_\Lambda$.
We refer to the latter as ``linear $\theta_\Lambda$'', in contrast to our method ``RBF $\theta_\Lambda$''.
\reftab{EnergyFitAblation} compares fitting error in both cases, with the same number of parameters (300).
It also compares results when fitting either energy or stress.
Note that energy fit with linear $\theta_\Lambda$ (equivalent to~\cite{sperl2020hylc}) suffers a stress error of more than $95\%$.
This error goes down to $25\%$ with our approach.

We have also validated the impact of stress normalization.
Under a common model complexity of 300 RBFs for all microstructures, the averaged error per stress component ranges between $7.2 \%$ and $62.7 \%$ without normalization, and $11 \%$ and $40.3 \%$ with normalization.
The lowest error is for principal stretch and the highest for principal curvature in both cases.
Normalization slightly reduces accuracy on stretch, but it largely improves accuracy on bending.


\subsection{Model Estimation Results}

We have fitted our model to a relative stress RMSE threshold of $10\%$ on all 10 microstructures.
\reftab{Fitting} summarizes the number of RBFs used and the final error.

For 3 of the microstructures, we could not reach an error of $10\%$, as the optimization process stagnated.
All these 3 microstructures present auxetic behavior and a tendency to buckle.
We believe that the difficulty of the optimization comes from discontinuities in the stress values, due to buckling.
Nevertheless, in all cases the final fitting error was below $16\%$.

\subsection{Controlled Validation Tests}

We have tested simulations with non-uniform strain, with exactly the same boundary conditions for fine simulations at micro-structure scale, as well as coarse simulations on triangle meshes.
We have used testing conditions inspired by stretch and pear-loop experiments used, respectively, for in-plane and out-of-plane cloth characterization~\cite{sperl2022eylsmpf}, as well as twist experiments that produce, at the same time, stretch strain and biaxial bending strain.
Furthermore, we have executed these experiments for material patches cut along 4 different directions for each micro-structure: $0^\circ$, $45^\circ$, $90^\circ$, and $135^\circ$.

\reffig{bigtable1} to \reffig{bigtable10} show the final simulation results of both fine (green) and coarse (purple) simulations, for all 10 microstructures, along all 4 directions.
For the stretch simulations, all cases show an excellent match.
It is worth noting that our fitted models reproduce the auxetic response of the micro-structures, and even complex effects such as the longitudinal folds shown by Microstructure 4.
For the bending simulations, the resulting accuracy is not as high, but the anisotropy effects are matched qualitatively, and differences across microstructures are evident.
For the twist simulations, the main folds and shapes are matched, but finer wrinkles are often missed.
We believe these are micro-scale effects, often due to buckling, that cannot be captured by a homogenized material.

We have compared quantitatively the bending tests vs. an energy-fit model~\cite{sperl2020hylc}.
With our model, the average error in the pear-loop ratio height-width ratio is $15 \%$ across all materials and directions, and the maximum error is $28.5 \%$.
With the energy-fit model, the average error is $21.5 \%$ and the maximum error is $106 \%$.
This result supports the higher accuracy of bending stiffness with our model.

Importantly, our coarse simulation results are obtained with meshes with just a fraction of the nodes of the fine simulations.
\reftab{Fitting} lists the mesh complexity of the fine simulations.
For the coarse simulations, we have used just 400 nodes for stretch and 900 for twist, for all materials.
The ratio fine vs. coarse ranges between $6$ (microstructure 5, twist) and $142$ (microstructure 7, stretch).
For the bending experiments we only simulated a thin strip, therefore the differences are not as dramatic.

\subsection{Real-World Validation Tests}

We have also tested simulations vs. real-world deformations with 3D-printed microstructures.
\reffig{comparisons} shows twist deformations on 4 microstructures.
We can see clear differences in the folding shapes across the 4 microstructures, and these differences are captured by our fitted materials.
The bottom row of the figure shows the result of an energy-fit model.
Some materials wrinkle too much, indicating that their bending stiffness is too low, which matches the result observed in the controlled validation tests.

%% file: includes/conclusion.tex



\section{Conclusions and Future Work}
\label{sec:concl}

We have presented a method for the homogenization of thin-shell microstructures, based on a novel type of RBF interpolant with attractive properties: it ensures model conservativeness by construction, it provides degrees of freedom to naturally fit stress data, and it ensures smoothness in polar coordinates that match well the domain of training data.
We have shown the qualitative match between the resulting homogenized material models and both simulated and 3D-printed microstructures. 
We have also shown fitting accuracy that is superior to previous methods.

The eigen-strain parameterization enjoys important benefits, but it also comes with numerical challenges.
We have introduced {\em polar difference} as a robust and elegant way of handling the periodicity of the parameterization.
And we propose blending of isotropic model evaluations to ensure differentiability in the isotropic case.
A more efficient treatment would be desirable for the latter challenge, but we are not aware that such treatment exists.

Our current solution suffers some limitations that could motivate further work.
One limitation is that the training process only explores uniaxial bending data.
While the quality of our biaxial bending validations based on twist experiments is high, during training there is no guarantee that the resulting model will be accurate, as there is no biaxial training data.
Unfortunately, it is not possible to obtain biaxial bending data from simulations with periodic boundary conditions.
Nevertheless, we think that the current model could serve as warm start for an additional training step using non-uniform deformation data.

Another possible limitation is the high number of RBFs in the resulting model.
Note that this is due to the high dimensionality of the function space, as the number of RBFs listed in \reftab{Fitting} (ranging $130$ to $500$) is comparable to extremely coarse regular grids ($3^5 = 243$, $4^5 = 1024$).
As an extension to our work, it might be sensible to balance high-dimensional and low-dimensional fitting.
One could start from low-dimensional energy terms designed in an additive manner~\cite{Miguel2016,sperl2020hylc}, and add local high-dimensional RBF corrections.
If the RBF radii are small, then distant RBFs can be pruned at runtime, reducing the actual number of RBF evaluations.

The final important limitation is that our model does not account for discontinuities or multi-stability.
We have confirmed that these phenomena occur in practice, due to microstructure buckling, and they hinder the fitting accuracy of our method for some materials.
The nonlinear activation functions in neural networks might be a natural answer to model discontinuities.
We think that imposing conservativeness is anyway a desirable property, and non-conservative effects due to, e.g., internal friction from microstructure contact, could be handled using a homogenized friction model on top of the pure conservative model~\cite{Miguel2013}.

To conclude, we would like to mention that our homogenized simulation renders were generated using barycentric interpolation of fine-scale geometry within the coarse triangle mesh.
This can be improved with state-of-the-art solutions for strain-dependent micro-scale rendering~\cite{Sperl2021,Montazeri2021}.

%% file: includes/validations.tex

\newcommand{\comparison}[1]{
\includegraphics[width =0.23\linewidth]{images/teaser/#1.PNG}
}

\newcommand{\comparisonRow}[1]{
\comparison{#11}
\comparison{#13}
\comparison{#14}
\comparison{#15}
}

\begin{figure*}[t]
\centering
\comparisonRow{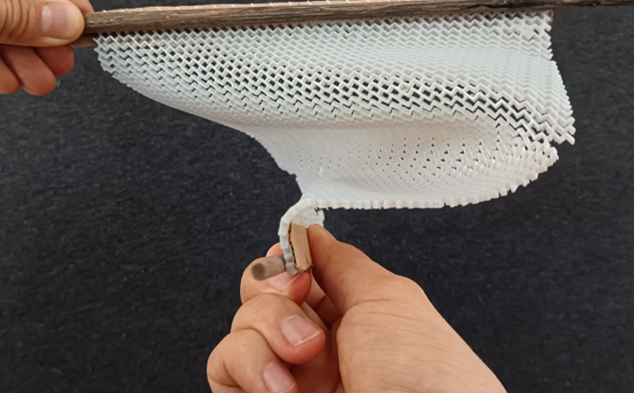}\\
\comparisonRow{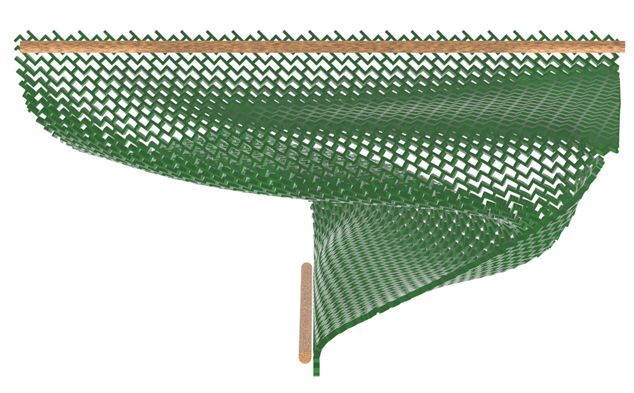}\\
\comparisonRow{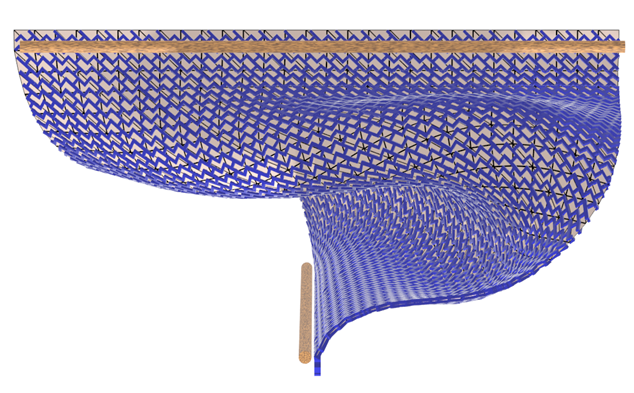}\\
\comparisonRow{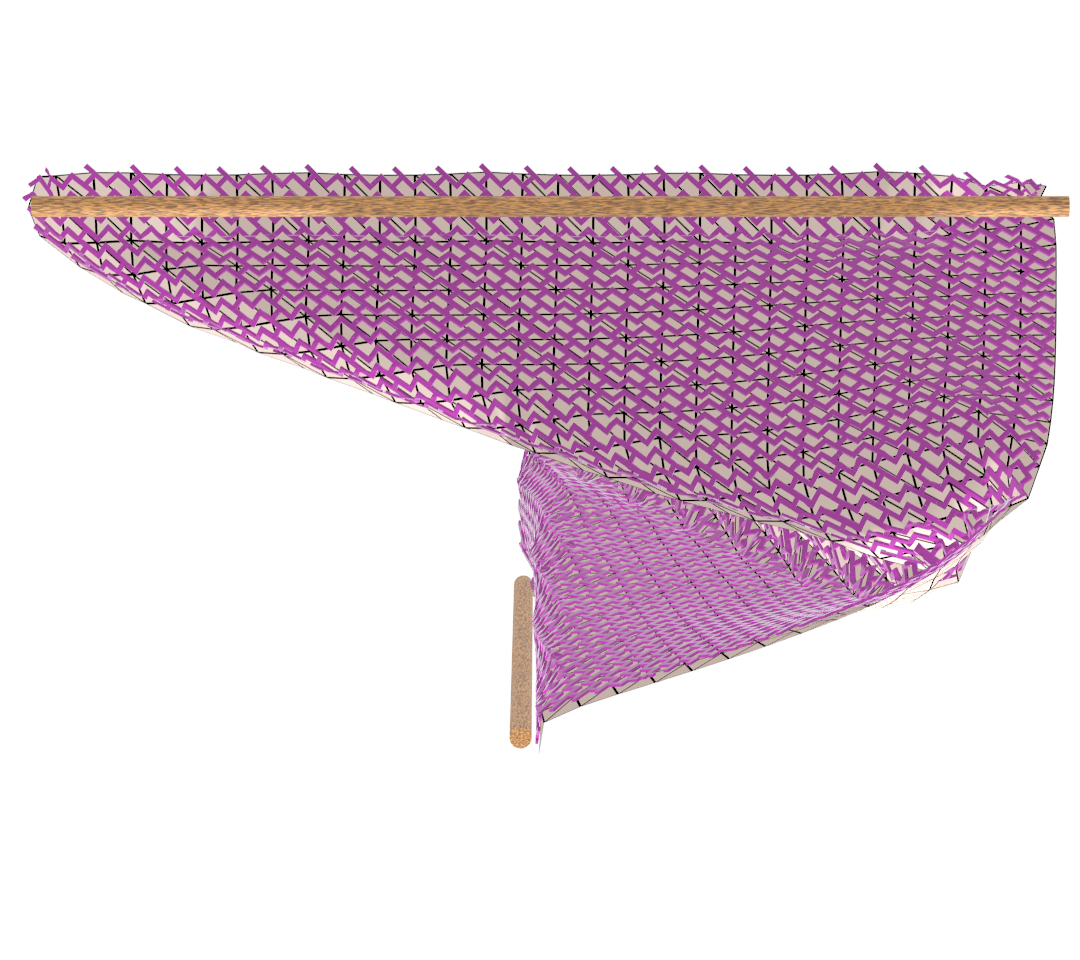}\\
\caption{
\label{fig:comparisons}
Same as in \reffig{teaser}, we compare twist deformations of real-world microstructures (top row), micro-scale simulations (second row), and macro-scale simulations computed with our homogenized material models (third row).
But we also add simulation results with an energy-fit model (bottom row).
From left to right, the images show microstructures 1, 3, 4 and 5.}
\Description[Comparison of simulation and real-world twists]{Comparison of real-world, micro-structure simulation, and homogenized simulation.}
\end{figure*}


\clearpage
\clearpage

\newcommand\meshRes{20}
\newcommand\coefType{6}
\newcommand\widthFraction{0.3}

\newcommand{\orientation}[2]{
\includegraphics[width=\widthFraction\linewidth]{images/MS_mesh\meshRes_coef\coefType_angle#2/output#1_stretch3_#2.jpg}
\includegraphics[width=\widthFraction\linewidth]{images/MS_mesh\meshRes_coef\coefType_angle#2/output#1_twist_#2.jpg}
\includegraphics[width=\widthFraction\linewidth]{images/MS_mesh\meshRes_coef\coefType_angle#2/output#1_pearloop_#2.jpg}\\
}

\newcommand{\microstructureBIS}[1]{
\begin{figure}
\centering
\orientation{#1}{0}
\orientation{#1}{45}
\orientation{#1}{90}
\orientation{#1}{135}
    \caption{Micro. #1 $50\%$ stretch, $90^\circ$ twist, pear loop ($0^\circ$, $45^\circ$, $90^\circ$, $135^\circ$) \label{fig:bigtable#1}}
\Description[Stretch, twist and bending for microstructure #1]{Comparison of stretch, twist and bending for microstructure #1, between microstructure and homogenized simulation.}
\end{figure}
}

\newcommand{\microstructure}[1]{
\begin{figure}[H]
\centering
\orientation{#1}{0}
\orientation{#1}{45}
\orientation{#1}{90}
\orientation{#1}{135}
    \caption{Micro. #1 $50\%$ stretch, $90^\circ$ twist, pear loop ($0^\circ$, $45^\circ$, $90^\circ$, $135^\circ$) \label{fig:bigtable#1}}
\Description[Stretch, twist and bending for microstructure #1]{Comparison of stretch, twist and bending for microstructure #1, between microstructure and homogenized simulation.}
\end{figure}
}








\begin{figure}[H]
\centering
\orientation{1}{0}
\orientation{1}{45}
\orientation{1}{90}
\orientation{1}{135}
    \caption{Micro. 1 $50\%$ stretch, $90^\circ$ twist, pear loop ($0^\circ$, $45^\circ$, $90^\circ$, $135^\circ$) 
    \label{fig:bigtable1}}
\Description[Stretch, twist and bending for microstructure 1]{Comparison of stretch, twist and bending for microstructure 1, between microstructure and homogenized simulation.}
\end{figure}

\begin{figure}[H]
\centering
\vspace*{-2\baselineskip}
\orientation{3}{0}
\orientation{3}{45}
\orientation{3}{90}
\orientation{3}{135}

    \caption{Micro. 3 $50\%$ stretch, $90^\circ$ twist, pear loop ($0^\circ$, $45^\circ$, $90^\circ$, $135^\circ$) 
    \label{fig:bigtable3}}
\Description[Stretch, twist and bending for microstructure 3]{Comparison of stretch, twist and bending for microstructure 3, between microstructure and homogenized simulation.}
\end{figure}

\begin{figure}[H]
\centering
\orientation{4}{0}
\orientation{4}{45}
\orientation{4}{90}
\orientation{4}{135}
    \caption{Micro. 4 $50\%$ stretch, $90^\circ$ twist, pear loop ($0^\circ$, $45^\circ$, $90^\circ$, $135^\circ$) 
    \label{fig:bigtable4}}
\Description[Stretch, twist and bending for microstructure 4]{Comparison of stretch, twist and bending for microstructure 4, between microstructure and homogenized simulation.}
\end{figure}

\begin{figure}[H]
\centering
\orientation{5}{0}
\orientation{5}{45}
\orientation{5}{90}
\orientation{5}{135}
    \caption{Micro. 5 $50\%$ stretch, $90^\circ$ twist, pear loop ($0^\circ$, $45^\circ$, $90^\circ$, $135^\circ$) 
    \label{fig:bigtable5}}
\Description[Stretch, twist and bending for microstructure 5]{Comparison of stretch, twist and bending for microstructure 5, between microstructure and homogenized simulation.}
\end{figure}

\begin{figure}[H]
\centering
\orientation{6}{0}
\orientation{6}{45}
\orientation{6}{90}
\orientation{6}{135}
\vspace*{-\baselineskip}
    \caption{Micro. 6 $50\%$ stretch, $90^\circ$ twist, pear loop ($0^\circ$, $45^\circ$, $90^\circ$, $135^\circ$) 
    \label{fig:bigtable6}}
\Description[Stretch, twist and bending for microstructure 6]{Comparison of stretch, twist and bending for microstructure 6, between microstructure and homogenized simulation.}
\end{figure}

\begin{figure}[H]
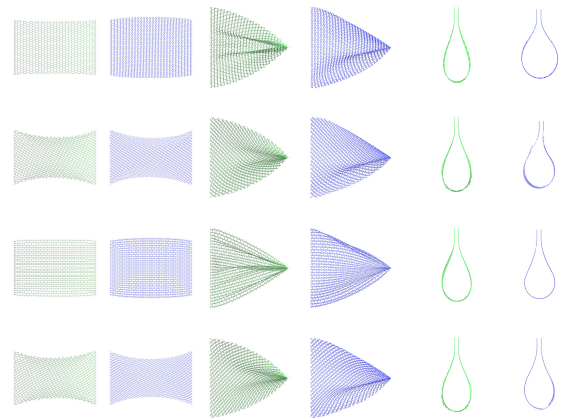

\centering
\orientation{10}{0}
\orientation{10}{45}
\orientation{10}{90}
\orientation{10}{135}
    \caption{Micro. 10 $50\%$ stretch, $90^\circ$ twist, pear loop ($0^\circ$, $45^\circ$, $90^\circ$, $135^\circ$) 
    \label{fig:bigtable10}}
\Description[Stretch, twist and bending for microstructure 10]{Comparison of stretch, twist and bending for microstructure 10, between microstructure and homogenized simulation.}
\end{figure}